\documentclass[a4paper,11pt]{article}
\pdfoutput=1 

\usepackage{jcappub} 

\usepackage[T1]{fontenc} 
\usepackage{hyperref}
\usepackage[dvipsnames]{xcolor}

\usepackage{cleveref}

\usepackage{slashed}
\usepackage{mathtools}
\usepackage{amssymb}
\usepackage{physics}
\usepackage{soul}

\newcommand{\DNeff}{\Delta N_{\rm eff}}
\newcommand{\be}{\begin{equation}}
\newcommand{\ee}{\end{equation}}
\newcommand{\bea}{\begin{eqnarray}}
\newcommand{\eea}{\end{eqnarray}}
\newcommand{\ba}{\begin{aligned}}
\newcommand{\ea}{\end{aligned}}

\title{Extending preferred axion models via heavy-quark induced early matter domination}

\author[a]{Andrew Cheek,}
\author[a]{Jacek K. Osi\'nski,}
\author[a,b]{Leszek Roszkowski}

\affiliation[a]{Astrocent, Nicolaus Copernicus Astronomical Center Polish Academy of Sciences, ul.~Rektorska 4, 00-614, Warsaw, Poland}
\affiliation[b]{National Centre for Nuclear Research, ul.~Pasteura 7, 02-093 Warsaw, Poland}

\emailAdd{acheek@camk.edu.pl}
\emailAdd{josin@camk.edu.pl}
\emailAdd{leszek.roszkowski@ncbj.gov.pl}

\abstract{We examine the cosmological consequences of the heavy quarks in KSVZ-type axion models. We find that their presence often causes an early matter domination phase, altering the evolution of the Universe. This extends the axion mass into the region where standard cosmology leads to overproduction, and allows for a greater number of axion models with non-renormalizable terms to be viable. Quantitatively, we find that decays proceeding through effective terms of up to dimension 9 ($d=9$) remain consistent with cosmological constraints, in contrast with the result $d\leq5$ previously found in the literature. As a consequence, the heavy quarks can be much heavier and the axion mass window with the correct relic density for dark matter is extended by orders of magnitude, down to $m_a\approx 6\times 10^{-9} \,{\rm eV}$. This is achieved without resorting to fine-tuning of the initial misalignment angle, bolstering the motivation for many future axion haloscope experiments. Additionally, we explore how these models can be probed through measurements of the number of relativistic degrees of freedom at recombination.
}

\begin{document}
\maketitle
\flushbottom

\section{Introduction}

The dark matter problem in modern physics is one with a great number of candidate solutions. Among these candidates, the QCD axion has emerged as a compelling contender, offering a remarkable explanation to two of the biggest challenges to fundamental physics: the dark matter, and the strong CP problems. The solution to the latter can be achieved by the Peccei-Quinn (PQ) mechanism~\cite{Peccei:1977hh, Peccei:1977np}, where a spontaneously broken, global Abelian $U(1)_{\rm PQ}$ symmetry is added to the Standard Model. After QCD confinement, the Nambu-Goldstone boson of the broken $U(1)_{\rm PQ}$, the axion, obtains a mass
\begin{equation}
    m_a = 5.7\times10^{-5}\,{\rm eV} \left(\frac{10^{11}\,{\rm GeV}}{f_a}\right).
    \label{eq:QCDaxionmass}
\end{equation}
The decay constant $f_a$ is related to the vacuum expectation value of the scalar field associated with PQ symmetry, $\langle \phi\rangle=f_a/\sqrt{2}$. Flavour-violating processes such as rare kaon decays require that $f_a$ is much larger than the electroweak scale, thus making viable QCD axions very light and hard to detect. For this reason, \textit{invisible axion models} are appealing as dark matter candidates. The attractiveness of axions as a dual solution to the strong CP and the dark matter problems is intensified when one realises that the cosmological evolution of the axion field leads to a cold matter relic abundance in the early Universe~\cite{Preskill:1982cy,Dine:1982ah, Abbott:1982af}.

Under Occam's razor we are compelled to consider this scenario as preferred until further evidence disfavors it. Additionally, the solution is highly predictive, motivating a huge experimental effort to test the interaction of axions with Standard Model (SM) particles, most notably the axion-photon coupling, $g_{a\gamma}$. Astrophysical objects provide a powerful tool for constraining the axion mass window~\cite{Raffelt:1990yz, Raffelt:2006cw}, but for the mass region most relevant for axion dark matter, around $10\,{\rm \mu eV}$, laboratory-based experiments are most promising~\cite{DiLuzio:2020wdo}. In fact, haloscope experiments~\cite{Sikivie:1983ip}, such as ADMX~\cite{ADMX:2009iij,ADMX:2019uok} and HAYSTAC~\cite{HAYSTAC:2018rwy} were able to reach the QCD axion band for a portion of the masses most relevant for dark matter. This is only the beginning and there are several operating and proposed haloscopes that are working to explore a much broader mass region, see Refs.~\cite{Irastorza:2018dyq, Sikivie:2020zpn, Billard:2021uyg, Adams:2022pbo} for reviews.

Within this picture however, it is important to take some deeper theoretical considerations into account, such as the UV origin of the colour anomaly, leading to additional particle content. Inspecting the effects of the additional particles led to the proposal of the \textit{preferred axion models}~\cite{DiLuzio:2016sbl,DiLuzio:2017pfr}, where some models were elevated by virtue of their consistency with the current standard cosmology and behaviour up to the Planck scale, $M_{\rm Pl}$. 

The cosmological arguments made for the preferred axion models largely rest on models in which one introduces new heavy quarks, $Q$, charged under $U(1)_{\rm PQ}$. In the post-inflationary scenario $Q$s will be produced through thermal freeze-out and to align with observations, should be unstable. This leads to the requirement that Lagrangian terms enabling $Q$ decay must proceed at a renormalizable level ($d\leq4$) or at the non-renormalizable level at most through dimension-5 ($d\leq5$) operators~\cite{DiLuzio:2016sbl,DiLuzio:2017pfr}.

In this paper we challenge this conclusion, showing that without any additional particle content, models where $Q$ decays to the SM at the $d>5$ are indeed allowed by existing cosmological constraints. This is because the new heavy quarks, although sufficiently unstable, in many cases will live long enough to dominate the cosmic energy budget altering the evolution of the Universe and, as a result, the dynamics of the misalignment mechanism~\cite{Steinhardt:1983ia, Lazarides:1990xp, Kawasaki:1995vt,Ramberg:2019dgi, Arias:2021rer,Arias:2022qjt}. This has the consequence of increasing the viable masses for the heavy quarks, as well as reopening the possibility for axion models that were previously disregarded. Importantly for axion dark matter, this provides a self consistent way to extend the axion mass to much lower values without invoking additional fields or fine-tuning in a pre-inflationary symmetry breaking scenario. 

Additionally, we investigate the impact of the decay of these heavy quarks into axions and how measurements of the number of relativistic degrees of freedom, $N_{\rm eff}$, can help constrain their branching ratios. This provides another avenue in which we alter the view on preferred axion models, since some lead to large branching ratios into axions and others not. Interestingly, we find that constraints are of a similar order, independent of whether the $Q$ particles dominate early on. 

This article is organised as follows, in section \ref{sec:Preferred_ams} we briefly review the relevant aspects of the preferred axion model framework. In section \ref{sec:EQD} we argue with approximations why and where we should expect early matter domination (EMD), then we support our claims with a fully fledged numerical treatment, showing that $Q$ decays of $d\geq6$ are completely consistent with observation. In section~\ref{sec:DNEFF} we explore the consequences of heavy quark decay for the additional amount of relativistic degrees of freedom, $\DNeff$. We do this for all decays, including the standard $d\leq5$ cases. Finally, in section~\ref{sec:conclusion} we conclude the article, emphasising future directions of study and implications on the search for axion dark matter.

\section{Preferred axion models}
\label{sec:Preferred_ams}

The associated Noether current $J_\mu^{\rm PQ}$ for the Abelian $U(1)_{\rm PQ}$ is conserved up to anomalies 
\begin{equation}
    \partial^\mu J_\mu^{\rm PQ} = \frac{g_s^2 N}{16\pi^2}G\tilde{G} + \frac{e^2 E}{16\pi^2}F\tilde{F},
\end{equation}
where the first term on the right is the colour anomaly, required for solving the strong CP problem, the second term is the electromagnetic anomaly, $G$ and $F$ are the field strength tensors of the gluon and electromagnetic fields , respectively, and $\tilde{G}$ and $\tilde{F}$ are their duals. The coefficients $g_s$ and $e$ are the Standard Model (SM) coupling constants, whereas $N$ and $E$ are the model-dependent anomaly coefficients. 

There are two classes of phenomenological axion models that generate a colour anomaly ~\cite{Kim:2008hd, DiLuzio:2020wdo}. The Kim–Shifman–Vainshtein–Zakharov (KSVZ) models, where SM fields are uncharged under $U(1)_{\rm PQ}$, introduce the colour anomaly via a heavy fermion, $Q$, which has colour charge~\cite{Kim:1979if,Shifman:1979if}. Alternatively, the Dine–Fischler–Srednicki–Zhitnisky (DFSZ) models imbue SM quarks with $U(1)_{\rm PQ}$ charges and introduce an additional weak doublet~\cite{Zhitnitsky:1980tq,Dine:1981rt}. The \textit{preferred axion model} framework~\cite{DiLuzio:2016sbl,DiLuzio:2017pfr} is strongly motivated by the cosmological implications of the heavy quarks and post-inflationary $U(1)_{\rm PQ}$ breaking such that the reheating temperature is greater than the breaking scale, $T_R>f_a$. Therefore, in this article we focus on the KSVZ setup, which supplements the SM Lagrangian, $\mathcal{L}_{\rm SM}$ by,
\begin{equation}
    \mathcal{L}_a = \mathcal{L}_{\rm SM} + \mathcal{L}_{\rm PQ} -V_{H\Phi} + \mathcal{L}_{\rm Qq}\, ,
\end{equation}
where 
\begin{equation}
    \mathcal{L}_{\rm PQ}= \vert\partial_\mu\Phi\vert^2 +\overline{Q}i\slashed{D}Q - (y_Q\overline{Q}_L Q_R \Phi + {\rm H.c})\,.
    \label{eq:LagPQ}
\end{equation}
The potential $V_{H\Phi}$ contains the new scalar couplings and $\mathcal{L}_{\rm Qq}$ contains terms that couple $Q$ to SM quarks, $q=q_L, d_R, u_R$. We see in eq.~\eqref{eq:LagPQ} that the Yukawa term will give rise to a mass term after PQ breaking, $m_Q=y_Q f_a$. For the purposes of this work, we take $y_Q=1$. Since by construction, $Q_L$ and $Q_R$ are non-trivially charged under $SU(3)_c$, they will strongly interact, annihilating into quarks and hadrons. In the early Universe, at temperatures above the QCD phase transition, $T_{\rm QCD}\sim 150\,{\rm MeV}$, annihilation can be perturbatively calculated for free $Q$s and SM quarks, 
\begin{equation}
    \langle\sigma v\rangle_{\bar{Q}Q}=\frac{\pi\alpha_s^2}{16 m_Q^2}\left(c_f n_f + c_g\right),
    \label{eq:sigmav}
\end{equation}
where $n_f$ is the number of quark flavours that $Q$ can annihilate into, and $\left(c_f, c_g\right)=\left(2/9,\,220/27\right)$ for triplets and $\left(c_f, c_g\right)=\left(2/3,\,27/4\right)$ for octets. With $T_R>f_a$, this interaction is strong enough to bring even very heavy $Q$s into thermal equilibrium with the SM. As with any massive particle, when the SM bath cools to $T\sim m_Q/25$, Boltzmann suppression starts to dominate and the particle \textit{freezes out}, leaving a relic abundance. In the case where $Q$ is stable or quasi-stable substantial cosmological issues are present, such as overabundance or disruption of Big Bang Nucleosynthesis (BBN)~\cite{Kawasaki:2004qu, Jedamzik:2006xz, Jedamzik:2007qk, Kawasaki:2017bqm}. 

Neither $\mathcal{L}_{\rm PQ}$ nor the scalar potential $V_{H\Phi}$ contain terms that allow for $Q$ decay, so a stable heavy quark is worth examining, as indeed has been done in the literature~\cite{DeLuca:2018mzn,Gross:2018zha}. However, the strong interaction rate of \cref{eq:sigmav} often overproduces $Q$ in the Universe such that $\Omega_Q>\Omega_{\rm DM}$, requiring that $m_Q\lesssim 10\,{\rm TeV}$~\cite{DiLuzio:2016sbl,DiLuzio:2017pfr}. This situation is potentially alleviated by noting that at temperatures below $T_{\rm QCD}$ hadrons start to form and non-perturbative effects dominate the behaviour of strongly-charged objects. If the heavy quarks are still prevalent in the Universe, exotic hadrons containing $Q$s may form, which will likely enhance the annihilation cross-section thanks to finite size effects. This could potentially restart $Q$-hadron annihilations, altering the heavy number density, see Refs.~\cite{Dover:1979sn,Griest:1989wd,Nardi:1990ku,Arvanitaki:2005fa,Kang:2006yd,Jacoby:2007nw,Kusakabe:2011hk,Gross:2018zha}. Despite this, the existence of heavy exotic hadrons is still experimentally disfavored because of a number of phenomenological consequences such as; the production of anomalously heavy isotopes~\cite{Perl:2001xi}, stellar and neutron star stability~\cite{Hertzberg:2016jie, Gould:1989gw}, as well as producing anomalously large heat flow in the Earth's interior~\cite{Mack:2007xj}. 

Therefore, the heavy quarks should be unstable~\cite{DiLuzio:2016sbl,DiLuzio:2017pfr}. The strongest constraints on unstable particles, which at some point where in thermal equilibrium, typically come from BBN. The limit of which is often taken in terms of the lifetime of the unstable particle, $\tau\lesssim 0.01\,{\rm s}$~\cite{Kawasaki:2017bqm}. This limit relies on the assumption of radiation domination throughout the early Universe~\cite{Allahverdi:2020bys}. A more robust limit is to ensure that the SM plasma temperature at time of $Q$ decay is warmer than when BBN starts, \textit{i.e.} $T_{\rm decay}^Q\gtrsim T_{\rm BBN}\approx3\,{\rm MeV}$~\cite{Kawasaki:2000en,Hannestad:2004px,Ichikawa:2005vw,Ichikawa:2006vm,deSalas:2015glj,Hasegawa:2019jsa}. With this in mind, we look to terms with heavy $Q$s and SM $q$'s at both the renormalizable and non-renormalizable levels,  
\begin{equation}
    \mathcal{L}_{Qq} = \mathcal{L}_{Qq}^{d\leq 4}+ \mathcal{L}_{Qq}^{d>4}  = \mathcal{L}_{Qq}^{d\leq 4} +\frac{1}{\Lambda^{(d-4)}}\mathcal{O}^{d>4} + {\rm h.c.}\, .
    \label{eq:LEFF}
\end{equation}
Note that there are only a few charge configurations that lead to $\mathcal{L}_{Qq}^{d\leq 4}\neq 0$ and the fact that both the $U(1)_{\rm PQ}$ and $U(1)_Q$ are expected to be broken at a scale up to the Planck scale, non-renormalizable terms are also considered. In Refs.~\cite{DiLuzio:2016sbl, DiLuzio:2017pfr} the suppression scale was taken `conservatively' to be $\Lambda=M_{\rm Pl}=1.22\times10^{19}\,{\rm GeV}$, but we stress that this is unknown. \\ 

\noindent The next restriction on axion models comes from the energy density of the axion itself, which is produced via the misalignment mechanism, resulting in an abundance~\cite{Sikivie:2006ni,Marsh:2015xka,Arias:2021rer}
\begin{align}
    \Omega_a h^2\approx
    \begin{dcases}
        0.006 \left(\frac{\theta_{\rm i}}{1}\right)^{2}\left(\frac{5.6~\mu {\rm eV}}{m_a}\right)^{3/2} &\text{ for } m_a \lesssim 3\, H(T_{\rm QCD})\,,\\
        0.17 \left(\frac{\theta_{\rm i}}{1}\right)^{2}\left(\frac{5.6~\mu {\rm eV}}{m_a}\right)^{7/6} &\text{ for } m_a \gtrsim 3\, H(T_{\rm QCD}), 
    \end{dcases}
    \label{eq:relic_std}
\end{align}
with the standard cosmological evolution. Here $m_a$ is the zero-temperature axion mass, $\theta_i$ is the initial angle of the axion potential and $T_{\rm QCD}$ is the temperature of the QCD phase transition. 

In the post-inflationary PQ symmetry breaking picture, different patches of the Universe randomly take $\theta_i$ values on the unit circle, which is then averaged to obtain $\langle\theta_i\rangle=\pi/\sqrt{3}$. By taking the thermal relic value from PLANCK, $\Omega_{\rm DM}=0.12$~\cite{Planck:2018vyg}, we can see that overabundance occurs for $m_a\lesssim 145\,{\rm \mu eV}$. Since $m_a$ is determined by the QCD anomaly \eqref{eq:QCDaxionmass}, we can read off an approximate limit for the breaking scale, $f_a\lesssim3\times 10^{11}\,{\rm GeV}$. By taking the maximum heavy quark mass $m_Q\sim f_a$, Ref.~\cite{DiLuzio:2017pfr} found that BBN constraints and $\Omega_a<\Omega_{\rm DM}$ requires that $Q$ decays must proceed through operators of dimension $d\leq 5$, thus providing a powerful limit on KSVZ type axion models, and a central criterion for \textit{preferred axion models}. 

This restriction on $d$ has the theoretically awkward consequence that the $U(1)_Q$ symmetry present in $\mathcal{L}_{\rm PQ}$ must be of a low-quality, \textit{i.e.} terms that break $U(1)_Q$ must occur at low dimension even with Planck-scale suppression. This is in contrast with the $U(1)_{\rm PQ}$ symmetry, which is required to be of high-quality, where terms that break PQ-symmetry can only appear at $d\geq 11$ in order not to spoil the solution to the strong CP problem~\cite{Kamionkowski:1992mf,Holman:1992us,Barr:1992qq}. Although this situation is not untenable and can be explained by additional discrete symmetries~\cite{Ringwald:2015dsf,DiLuzio:2016sbl}, it would be theoretically appealing to increase the quality of $U(1)_Q$. 

In this work we present our findings allowing the initial angle to vary. $\theta_i\in\left[0.5,\pi/\sqrt{3}\right]$, we do this to be more agnostic on the value of $\langle\theta_i\rangle$ and capture some of the effects that could be relevant for the pre-inflationary breaking scenario without resorting to a finely tuned angle of $\theta_i$, which is ultimately limited by isocurvature fluctuations~\cite{Hertzberg:2008wr}. 

Another consequence of the post-inflationary PQ breaking is the expected formation of topological defects such as cosmic strings and domain walls~\cite{Kibble:1976sj,Kibble:1982dd,Vilenkin:2000jqa}, both of which could contribute to the density of relic axions, however, the exact contribution is still largely in dispute due to the technical challenges when it comes to calculating the effect, see Refs.~\cite{Hagmann:2000ja,Wantz:2009it, Hiramatsu:2010yu, Kawasaki:2014sqa, Gorghetto:2018ocs, Buschmann:2021sdq}. Therefore we do not consider these contributions and stick simply to the misalignment mechanism. Furthermore, domain walls can be cosmologically dangerous due to their scaling behaviour~\cite{Sikivie:1982qv}. This issue is avoided by only considering models where the domain walls do not form, specifically this is when the domain wall number is one, $N_{\rm DW}=1$~\cite{Vilenkin:1982ks,Barr:1986hs}. However, since there are multiple ways around this problem~\cite{Sikivie:1982qv,Turner:1985si,Choi:1996fs}, it isn't considered a strong requirement. Despite this, it is worth noting that including this requirement with the finding of Refs.~\cite{DiLuzio:2016sbl,DiLuzio:2017pfr}, \textit{i.e.} that $Q$ decays must proceed through terms of dimension $d\leq 5$ one is left with two hadronic axion models~\cite{Alonso-Alvarez:2023wig}
\begin{equation}
    {\rm KSVZ-I}\,:\,\left(3,1,-1/3\right),\,\,\,\,\,{\rm or}\,\,\,\,\,{\rm KSVZ-II}\,:\,\left(3,1,+2/3\right),
\end{equation}
where the brackets represent the SM charges, $\left({\rm SU}(3)_c,\,{\rm SU}(2),\, {\rm U}(1)_Y\right)$. From here one is free to choose the PQ charges $\left(\chi_L, \chi_R\right)$ of the left-handed and right-handed heavy fermions, but this still is rather restrictive.

In the next section we will discuss how the unstable heavy fermions can affect the cosmological evolution, the misalignment mechanism, and thus the window of \textit{preferred axion models}.

\section{Early heavy Q domination and axion relic abundance}
\label{sec:EQD}

As described above, the axion models of interest here are those with very heavy $Q$ fermions that have a charge under ${\rm SU}(3)_c$. The fast interaction rate ensures that above the temperature $\sim m_Q$, the $Q$s will be in thermal equilibrium. Once the temperature of the plasma cools below $m_Q$ the heavy particle freezes out in the standard way described by the Boltzmann equation for the number density of $Q$, $n_Q$,
\be
\frac{\dd n_{Q}}{\dd t} + 3H n_{Q}=  -\langle\sigma\,v\rangle\left[n_Q^2 -(n_Q^{\rm eq})^2\right]\label{eq:BE_nQ},
\ee
where for now we ignore the decay terms of $Q$. The superscript ``${\rm eq}$'' denotes the equilibrium distribution, which gets Boltzmann suppressed at temperatures below the particle mass, $T<m$. To get an idea about the temperatures the thermal bath may have when a meta-stable $Q$ starts to dominate the energy of the Universe, we can take the approximate textbook result 
\begin{equation}
    Y_Q^{\rm f}=\frac{n_Q}{s_{R}^{\rm SM}}\approx \frac{45}{2\pi^2 g_{\star S}(T_{\rm f})} \frac{x_{\rm f}^3 H(T_{\rm f})}{m_Q^3\langle\sigma v\rangle}, 
\end{equation}
where $s_{R}^{\rm SM}$ is the SM plasma entropy density and \(T_{\rm f} = m_Q/x_{\rm f }\) is the freeze-out temperature and numerically \(x_{\rm f} \approx 25\). The effective number of entropy degrees of freedom is given by $g_{\star S}$. Assuming a radiation-dominated Universe prior to and at the time of freeze-out one can obtain an estimate for the ratio of energy densities for $Q$ and SM radiation
\be
\frac{\rho_Q}{\rho_{\rm R}^{\rm SM}}=\frac{m_{Q}\,n_{Q}\left(m_Q, T\right)}{\rho_{\rm R}^{\rm SM}\left(T\right)}\sim 10^{10}\left(\frac{m_Q}{10^{12}\,{\rm GeV}}\right)^2\left(\frac{1\,{\rm MeV}}{T}\right), \label{eq:est_MD}
\ee
where we have scaled $m_Q$ and $T$ values to the approximate mass scale for the misalignment mechanism to produce axion dark matter and the temperature at BBN ($T\sim 3\,{\rm MeV}$). This implies that the $Q$ particles of mass $10^{12}\,{\rm GeV }$ will start to dominate the Universe at $T_{\rm eq.}\sim 10^7\, {\rm GeV}$, well above the temperature at the onset of BBN, therefore, if the $Q$ decays are slow enough, many axion models in fact feature a period of EMD. 

To calculate the total decay width, $\Gamma_Q$, we approximate the behaviour of $\mathcal{L}_{Qq}$, following Ref.~\cite{DiLuzio:2017pfr} by assuming a constant matrix element and massless final states to obtain an analytic phase space integral~\cite{DiLuzio:2015oha},
\begin{equation}
    \Gamma_{d,n_f}=\frac{m_Q}{4\left(4\pi\right)^{2n_f-3}\left(n_f-1\right)!\left(n_f-2\right)!}\left(\frac{m_Q^2}{\Lambda^2}\right)^{d-4},
\end{equation}
where $n_f$ is the number of final state particles, for $d=5,6,7,8$ we have at least $n_f=2,3,4,5$. This leads to the corresponding decay widths,
\begin{align}
    \Gamma_{d=4} &= \frac{m_Q}{8\pi}, \,\,\,\, &\Gamma_{d=5}&= \frac{m_Q^3}{16\pi \Lambda^2},\nonumber\\ 
    \Gamma_{d=6}&= \frac{m_Q^5}{512\pi^3 \Lambda^4},\,\,\,\,  &\Gamma_{d=7}&= \frac{m_Q^7}{49152\pi^5 \Lambda^6}, \nonumber\\
    \Gamma_{d=8}&= \frac{m_Q^9}{576\left(4\pi\right)^7 \Lambda^8},\,\,\,\,  &\Gamma_{d=9}&= \frac{m_Q^{11}}{11520\left(4\pi\right)^9 \Lambda^{10}}\,, \,
    \label{eq:EFFwidths}
\end{align}
where, unlike Ref.~\cite{DiLuzio:2017pfr} we do not set $\Lambda\rightarrow M_{\rm Pl}$, although this is a reasonable upper limit to take because it is expected that the PQ symmetry will be broken at or below such a scale.

To estimate the plasma temperature at time of $Q$ decay, $T^Q_{\rm decay}$, we set $\Gamma_Q \equiv H_{\rm rad}\left(T=T^Q_{\rm decay}\right)$ where $H_{\rm rad}$ is the Hubble parameter in the radiation-dominated Universe.  
Generally this leads to the equation
\be
T^Q_{\rm decay}=\left(\frac{90M_{\rm Pl}^2}{8\pi^3}\frac{(\Gamma^Q_d)^2}{g_{\star}\left(T^Q_{\rm decay}\right)}\right)^{1/4}.
\ee
where numerically one has to iterate with the $T$ dependence in $g_\star$ to get $T^Q_{\rm decay}$. For $d=4$ the decay rate is very fast and typically $T^Q_{\rm decay}> T_{\rm f}$. However for dimension-5 decays, with maximum suppression $\Lambda=M_{\rm Pl}$, we have
\begin{equation}
    T^{Q,\,d=5}_{\rm decay} \approx 3.2\times 10^{7}\, {\rm GeV}\,\left(\frac{m_Q}{ 10^{12}\,{\rm GeV}}\right)^{3/2}\, \left(\frac{1}{g_\star\left(T^{Q,\,d=5}_{\rm decay}\right)}\right)^{1/4}.
\end{equation}
By inspecting when $T^{Q,\,d=5}_{\rm decay} \sim T_{\rm eq.}$, this equation suggests that $d=5$ decays could produce EMD for masses $m_Q\gtrsim 10^{12}\,{\rm GeV}$. However, by comparing $T_{\rm eq}$ and $T_{\rm decay}^{Q,\,d=5}$, we find that the period for $Q$-domination is very short. On the other hand, models with dimension $d\geq 6$ decays have a much greater time separation between the end of freeze-out, the temperature of quark domination and decay. This means the period of EMD is much longer and thus has a greater influence. 

\begin{figure}
    \centering
    \includegraphics{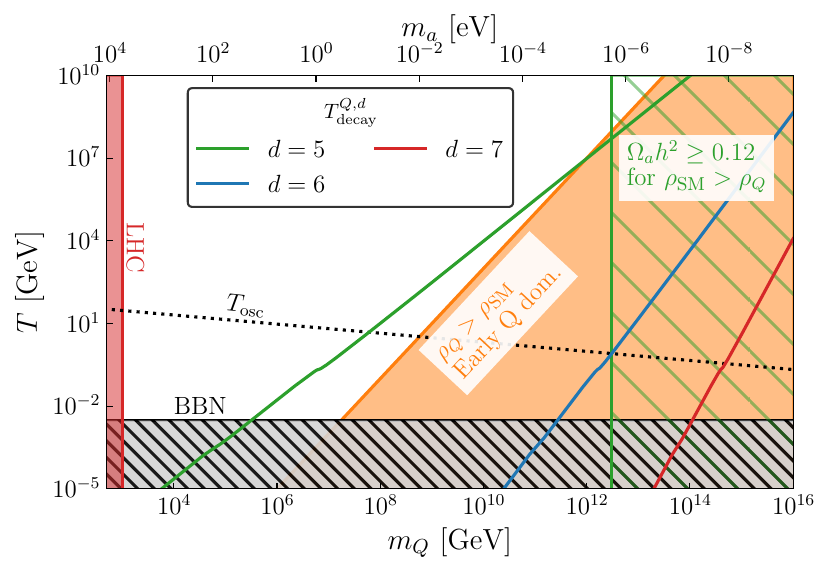}
    \caption{Heavy quark lifetimes shown in terms of the temperature of the SM plasma at time of decay $T_{\rm decay}^Q$. Different colored lines show the lifetimes via $d=5,\,6,\,7$ effective operators. LHC constraints are shaded red and BBN constraints from $T\lesssim3\,{\rm MeV}$ are hatched gray. The orange shaded region shows where our estimated early $Q$ domination should occur according to eq~\eqref{eq:est_MD}. The green hatched region is where the misalignment mechanism overproduces axions, $\Omega_a > \Omega_{\rm DM}$, assuming standard cosmology, if we take the $\theta_i=0.5$ value. The temperature at which the misalignment mechanism starts to produce particles, $T_{\rm osc}$, is shown with the black dotted line, also assuming standard cosmology. Therefore we expect variation to the green hatched region and black dotted line when they overlay with the orange region. The top axis is the axion mass given by eq.~\eqref{eq:QCDaxionmass} with $f_a=m_Q$.}
    \label{fig:diluca_similar}
\end{figure}

We sketch this out in Fig.~\ref{fig:diluca_similar} where we show the $Q$ lifetimes for different dimensions in terms of temperature at decay. Indeed our estimate for $\rho_{Q}\approx\rho_{\rm SM}$ intersects with $T^{Q,\,d=5}_{\rm decay}$ at around $m_Q\sim 10^{12}\,{\rm GeV}$ but the difference between $T_{\rm eq}$ and $T_{\rm decay}$ is not that large. We show where our above estimate predicts early $Q$ domination by the orange shaded region and we see that for $d=6$ \& $7$ this should occur for a substantial period of time. We also show the region of parameter space where misalignment would overproduce cold axions, assuming radiation domination and initial angle $\theta_i=0.5$\,\footnote{Ref.~\cite{DiLuzio:2017pfr} chose $\theta_i=\pi/\sqrt{3}$ to define a more aggressive constraint.}. Importantly this figure suggests that the assumption of standard early cosmology is not well founded for $d\geq 5$ for $m_Q$ above $\sim 10^{11}\,{\rm GeV}$.  

It has been shown that a period of EMD can augment the dynamics of the misalignment mechanism. The important physical quantity to evaluate when the axion field starts to oscillate around its true minimum is the oscillation temperature, $T_{\rm osc}$, defined by $3H\left(T_{\rm osc}\right)=\tilde{m}_a\left(T_{\rm osc}\right)$. The temperature-dependent axion mass $\tilde{m}_a(T)$ is given by
\begin{equation}
    \tilde{m}_a^2(T)=\frac{\chi(T)}{f_a^2}\, ,\label{eq:Tdepaxionmass}
\end{equation}
where $\chi(T)$ is the QCD susceptibility, which has been calculated using lattice QCD~\cite{Borsanyi:2016ksw}. In the WKB approximation, $\theta\ll 1$ and slowly varying $H$ and $\tilde{m}_a$, and a radiation-dominated early Universe, the resulting oscillation temperature is~\cite{Arias:2021rer}
\begin{align}
    T_{\rm osc}
    \begin{dcases}
        \left(\frac{1}{\pi}\sqrt{\frac{80 \pi}{g_{\star}\left( T_{\rm osc}\right)}}m_a M_{\rm Pl}\right)^{1/2} &\text{ for } T_{\rm osc} \leq  T_{\rm QCD}\,,\\
        \left(\frac{1}{\pi}\sqrt{\frac{80 \pi}{g_{\star}\left( T_{\rm osc}\right)}}m_a M_{\rm Pl} T_{\rm QCD}^4\right)^{1/6} &\text{ for } m_a \gtrsim 3\, H(T_{\rm QCD}). 
    \end{dcases}
    \label{eq:T_osc_std}
\end{align}
This estimate is depicted in Fig.~\ref{fig:diluca_similar} by a black dotted line. We can get an idea of whether early $Q$ domination is likely to affect the dynamics of misalignment by seeing if $T_{\rm osc}>T_{\rm decay}^Q$ inside the orange region. One can see that for $d=5$ decay occurs at temperatures far too high and thus the misalignment mechanism is unlikely to be altered, but for $d\geq 6$ we can expect some effects. Interestingly this will occur in the region which is prohibited by axion overproduction in standard cosmology. The effects of EMD have been shown to dilute the axion abundance if the matter decays after $T_{\rm osc}$~\cite{Steinhardt:1983ia, Lazarides:1990xp, Kawasaki:1995vt}. Therefore, this picture suggests that the axion models where $Q$ decays through $d\geq 6$ effective operators may in fact not be inconsistent and should not be disregarded. We evaluate this more precisely below.

\subsection{Full numerical treatment}
The estimates above only imply that $d\geq 6$ axion models should be reassessed by the preferred axion models framework. To properly determine whether this is the case we perform a full numerical evaluation of the coupled Friedmann-Boltzmann equations along with the axion equation of motion. These track the evolution of particle species that were produced thermally or through $Q$ decay, or via the misalignment mechanism. The Friedmann equation is
\be\label{eq:Hubble}
\frac{3H^2M_{\rm Pl}^2}{8\pi}=\rho_\mathrm{R}^{\rm SM} + \rho_{a} + \rho_{Q}\,,
\ee
where $\rho$ is the energy density, denoted for each component via the super/sub-scripts. The set of Boltzmann equations that describe the dynamics of the SM plasma entropy density, $s_{R}^{\rm SM}$, the $Q$ number density, $n_Q$, and the axion energy density, $\rho_{a}$, through particle creation/annihilation processes are
\begin{subequations}\label{eq:FBEqs}
\bea
 \frac{\dd s_\mathrm{R}^{\rm SM}}{\dd t} &=& -3H s_\mathrm{R}^{\rm SM}  +\frac{{\rm BR}_{\rm SM}\Gamma_Q}{T}\rho_{Q} \,,\\
 \frac{\dd\rho_a}{\dd t}  &=& -  4H \rho_a+{\rm BR}_{a}\Gamma_Q\rho_{Q}+\langle E_{\rm scat}^{a}\rangle\gamma_a\left(1-\frac{n_a}{n_a^{\rm eq}}\right)\,,\label{eq:FBEqs_rhoa}\\
 \frac{\dd n_{Q}}{\dd t} &=& - 3H n_{Q} -\Gamma_Q n_Q -\langle\sigma\,v\rangle\left[n_Q^2 -(n_Q^{\rm eq})^2\right] \,,
\eea
\end{subequations}
where the thermalized annihilation cross-section, $\langle\sigma v\rangle$, has been introduced in \cref{eq:sigmav}. The third term on the right hand side of eq.~(\ref{eq:FBEqs_rhoa}) determines whether the axions are in thermal equilibrium with the SM plasma, which proceeds via the thermal axion production rate $\gamma_a$~\cite{DEramo:2021lgb}. The averaged axion scattering energy, $\langle E_{\rm scat}^{a}\rangle$, describes the average energy released in each collision between the thermal bath and the axion, $\sim 3 T$~\cite{Giudice:2000ex}. We find that when heavy $Q$ domination occurs this term is irrelevant, but in section \ref{sec:DNEFF} we explore $d\leq5$ cases where it becomes important, and we discuss it in more detail there. We denote `simplified' branching ratios for $Q$ decay into SM and axions by ${\rm BR_{SM}}$ and ${\rm BR}_a$, respectively. These are simplified because we have performed our analysis in a model-independent way and focus primarily on the energy transfer between different species. For example, with the KSVZ-I model which has $\left(\chi_L, \chi_R\right)=(-1,-2)$, the lowest-order $Q$ decay term is
\be
\mathcal{L}_{Qq}= \Phi^\dagger \overline{Q}_{L}d_R +\frac{1}{\Lambda^{(d-4)}}\mathcal{O}^{d>4} 
\ee
therefore the conventional particle physics branching ratio for $Q\rightarrow a + d$ would be $1$, but because the energy will be distributed evenly to both the SM and axion energy densities, we will have ${\rm BR}_{\rm SM}={\rm BR}_a=1/2$. We have performed the calculation this way to preserve model independence and in section \ref{sec:DNEFF} we will elucidate implications for some specific models. In this section however we assume ${\rm BR}_{\rm SM}=1.0$ such that all axions are produced either thermally from bath particles or the misalignment mechanism.

To model the misalignment mechanism one solves the equation of motion for the axion field, $\tilde{a}$, which is expressed in terms of the axion angle, $\theta=\tilde{a}/f_a$

\begin{equation}\label{eq:misalignment}
    \left(\frac{\dd^2}{\dd t^2} + 3H(t)\frac{\dd}{\dd t}\right)\theta(t) + \tilde{m}_a^2(t)\sin(\theta(t))=0\, , 
\end{equation}
where $\tilde{m}_a(t)$ is the temperature- (and therefore time-) dependent mass of the axion defined in eq.~\eqref{eq:Tdepaxionmass}.

\begin{figure}[t!]
    \centering
    \includegraphics[width=\textwidth]{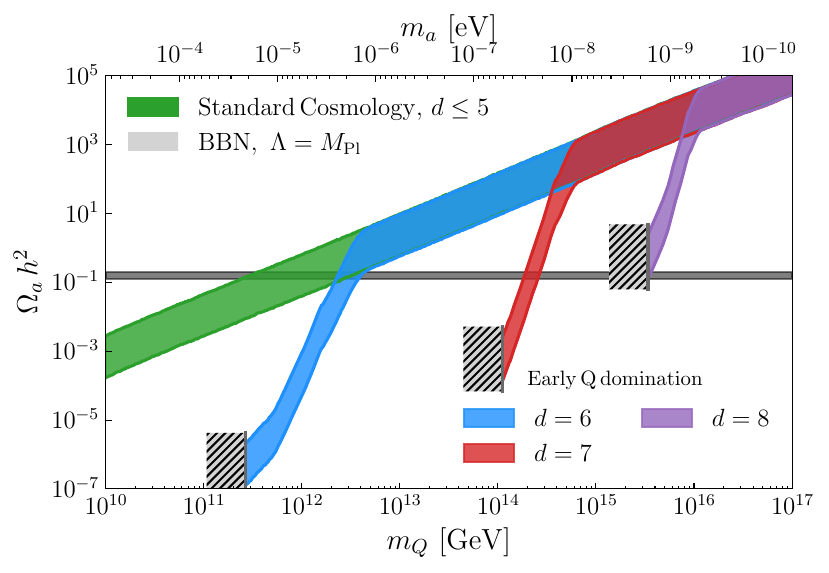}
\caption{Misalignment abundance of axions $\Omega_a$ where heavy $Q$s can decay via operators of different dimensions, $d\leq5\,({\rm green}),\,d=6\,({\rm blue}),\,d=7\,({\rm red}),\,{\rm and}\,d=8\,({\rm purple})\,$. The horizontal dark grey band shows the value for $\Omega_{\rm DM}$ as reported by the PLANCK collaboration~\cite{CMB-HD:2022bsz}.}
    \label{fig:misalignment_all_Ds}
\end{figure}

In practice, we solve the background cosmology described in equations \eqref{eq:Hubble} and \eqref{eq:FBEqs}, obtaining the resultant $H$ up until the time of BBN. We then solve eq.~\eqref{eq:misalignment} using the dedicated misalignment solver {\tt MiMeS}~\cite{Karamitros:2021nxi}. Indeed as our estimates suggested we observe significant regions of parameter space where heavy quarks dominate the early Universe and alter the cold relic axion abundance. In Fig.~\ref{fig:misalignment_all_Ds} we show the axion abundance for different dimensions in $\mathcal{L}_{Qq}$, $d=5,\,6,\,7,\,8$ in green, blue, red, and purple, respectively. The bands extend over the domain of initial misalignment angles which are not fine-tuned, $\theta_i\in\left[0.5,\pi/\sqrt{3}\right]$, showing how $\Omega_a$ varies over $m_Q=f_a$. Remember that the abundance $\Omega_a\propto \theta_i^2$ in the standard scenario so $\theta_i=\pi/\sqrt{3}$ gives the lower $m_Q$ in the bands shown. For this figure, we have set the cut-off scale for decays to $M_{\rm Pl}$ and convey where BBN constraints become relevant (\textit{i.e.} $T_{\rm decay}^{Q}\lesssim 3\,{\rm MeV}$) by black hatching.  

As anticipated by the approximations in Fig.~\ref{fig:diluca_similar}, our full calculations exhibited in Fig.~\ref{fig:misalignment_all_Ds} confirm that the misalignment mechanism is not altered by $d\leq5$ decays, and what we observe is the result which corresponds to the standard cosmology of radiation domination. However, $d=6,7,8$ do diverge from the standard misalignment mechanism, in the process varying the $f_a$ (and $m_a$) that corresponds to the correct relic abundance for axion dark matter. We see that for all cases, when $m_Q$ is high enough they will tend to the standard cosmology limit. This is consistent with Fig~\ref{fig:diluca_similar}, where the heavy quarks have decayed before the axion field starts to oscillate, $T_{\rm decay}^Q>T_{\rm osc}$. Of course $T_{\rm osc}$ is itself also affected by matter domination, but the rough picture still corresponds well. In both figures we see that $m_Q\gtrsim3\times 10^{12}\,{\rm GeV} \,{\rm and}\,\gtrsim7\times 10^{14}\,{\rm GeV}$ should correspond to the standard misalignment result for $d=6$ and $7$, respectively. 

By taking $\Lambda=M_{\rm Pl}$, and assuming standard cosmology, we see that $d\geq 6$ would come into tension with BBN constraints especially if you take the value $\theta_i=\pi/\sqrt{3}$. This is what is the primary motivator for limiting the preferred axion models to $d\leq 5$. However, when including the effects that frozen-out $Q$s will have on the early Universe the result is that axions are only overproduced when $m_Q\gtrsim2\times 10^{12}\,{\rm GeV}$. For $d=7$ overproduction occurs when $m_Q\gtrsim2\times 10^{14}\,{\rm GeV}$. 

As mentioned above we include the standard $\theta_i$ range to explore how reasonable variations on $\theta_i$ affect the results. We can see that for axion dark matter, dimension-6 and 7 models have a much narrower mass window than in the standard case. We even see that we can push the dimension of effective operators to 8, but BBN limits occur almost at the same point as $\Omega_a\approx \Omega_{\rm DM}$, making it impossible for $\theta_i=\pi/\sqrt{3}$ to obtain the correct relic abundance. However, the suppression scale $\Lambda$ is ultimately unknown. It is expected to break at most at the Planck scale, but potentially below it. 

\begin{figure}[t!]
    \centering
    \includegraphics[width=\textwidth]{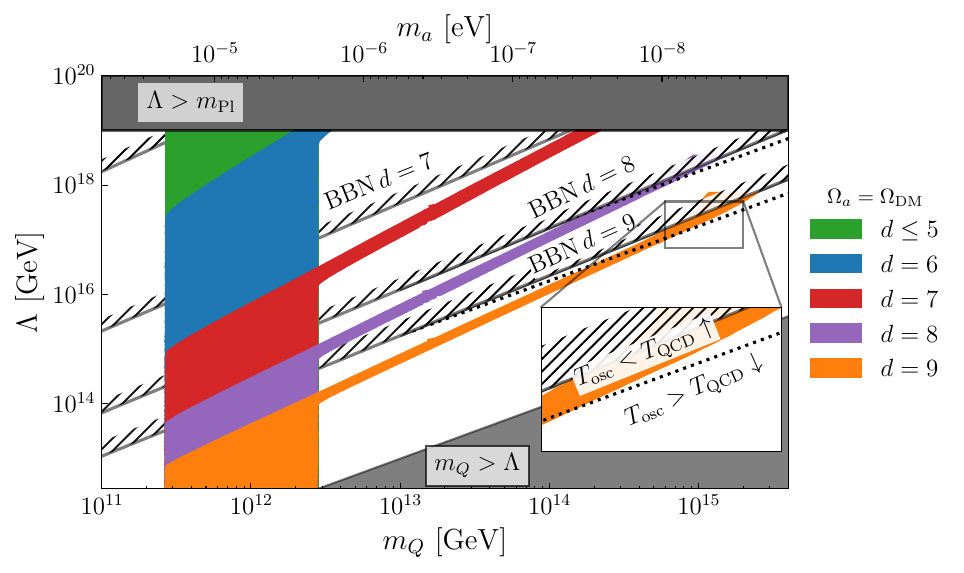}
    \caption{The values of $m_Q$ and $\Lambda$ that satisfy the correct relic abundance through the misalignment mechanism assuming an initial misalignment angle which is not fine-tuned, $\theta_i\in\left[0.5,\pi/\sqrt{3}\right]$ . Colours blue, red, purple, and orange are for $d=6,7,8\,$ and $9$, respectively. }
    \label{fig:scan_mQ_Lam}
\end{figure}

In Fig.~\ref{fig:scan_mQ_Lam} we treat $\Lambda$ as a free parameter. Now the shaded regions show what values of $m_Q$ and $\Lambda$ achieve the correct relic $\Omega_{\rm DM}$ with the domain $\theta_i\in\left[0.5,\pi/\sqrt{3}\right]$. We see that by varying $\Lambda$, decays at larger values of $d$ can occur without altering the misalignment mechanism. This is shown by the large vertical regions of each case where $m_a\sim\left[2\times 10^{-5}-2\times 10^{-4}\right]\,{\rm eV}$ corresponding to the standard-cosmology window. More interesting is when the coloured shaded regions diverge from this pattern, when EMD affects the misalignment mechanism. We see that the bands are much more narrow, and occur for larger values of $\Lambda$ and $m_{Q}$. This is because a larger $\Lambda$ suppresses the decay rate, making the $Q$-domination period longer. At the same time, increasing $m_Q$ increases $\Gamma_Q$ to a higher power than $\Lambda$. The driving effect for higher $m_Q$ is that their freeze-out is sooner and the corresponding yield is larger (see eq.~\eqref{eq:est_MD}), further contributing to an increased period of EMD. 

Once again the left and right boundaries of the shaded regions correspond to $\theta_i=\pi/\sqrt{3}$ and $0.5$, respectively. Above and to the left of each region corresponds to underproduction of axions with respect to $\Omega_{\rm DM}$ and therefore remains a viable area of parameter space for axion models where dark matter is not fully axionic. This is limited from above by the BBN constraints that are shown for different dimensions. Below and to the right of the shaded colored regions are areas that would require a fine-tuned $\theta_i<0.5$ to not overproduce the axions, helping to put limits on the \textit{preferred axion window}. We see also that by varying $\Lambda$ we can include $d=9$ and the dark matter abundance can be realised at values of $m_Q\sim 10^{15}\,{\rm GeV}$. As exhibited by the top axis in our figure, the variation in $m_Q$ implies that the axion mass can vary by orders of magnitude, bringing the benchmark region into the realms of low-mass axion experiments such as ADMX~\cite{Stern:2016bbw}, ABRACADABRA~\cite{Salemi:2021gck}, FLASH~\cite{Alesini:2017ifp}, DMRadio~\cite{DMRadio:2022pkf}, and SRF~\cite{Berlin:2020vrk}. This is the first time that the \textit{preferred axion models} are shown to actually predict masses down to $m_a\sim 10^{-8}\,{\rm eV}$ without invoking fine-tuning in $\theta_i$ or additional fields that alter the cosmology.

To determine the relation between $\Lambda$ and $m_Q$ that these early $Q$-dominated branches exhibit we refer to the expressions derived in Appendix B of Ref.~\cite{Arias:2021rer}. Our results are consistent with both the non-adiabatic and adiabatic EMD expressions when $T_{\rm osc}>T_{\rm QCD}$: $\Lambda \propto m_Q^{\frac{d-11/4}{d-4}}$ and $\Lambda \propto m_Q^{\frac{d-65/22}{d-4}}$, respectively. This leads us to an important point that was alluded to in the Introduction, namely the question of exotic hadron formation and how their formation could affect the physics of the early Universe. Typically the formation of these objects is considered in a radiation-dominated Universe and would be worth exploring in the context of $Q$ domination, but this is beyond the scope of this work. To estimate where exotic hadron formation may affect the misalignment mechanism we have determined where $T_{\rm osc}=T_{\rm QCD}$, shown as black dotted lines in Fig.~\ref{fig:scan_mQ_Lam}. There are two lines shown, one each for $d=8\,{\rm and}\,9$, because for the others, the line is in the region already constrained by BBN. For $d=9$ we show a zoomed in region to show that there is indeed a small region where $T_{\rm osc}<T_{\rm QCD}$, so that misalignment may occur during this period where exotic hadrons have formed. In this period of heavy $Q$ domination, the likely hadrons form into $\overline{Q}Q$ heavy mesons, thus continuing the period of EMD and therefore not altering the results shown in this section.

\section{Constraints from $\Delta N_{\rm eff}$}
\label{sec:DNEFF}
We have shown in the previous section that the heavy quarks required in KSVZ-type axion models may produce a period of EMD which can alter the dynamics of the misalignment mechanism. At the end of the early matter stage, the dominant energy density is required to transfer predominantly to SM radiation in order to fit with BBN and the formation of the Cosmic Microwave Background (CMB). For the latter, any additional radiation alters small-scale anisotropies and is parameterized by the effective number of relativistic degrees of freedom, $N_{\rm eff}$. This quantity is highly sensitive to the thermal history of particle physics content. At high temperatures, all SM particles, including neutrinos, are in thermal equilibrium. Through expansion and cooling, most particles thermally decouple when temperatures fall below their mass, $T<m$. Only neutrinos decouple when $T\gg m_{\nu}$, and later entropy injections to the SM bath separate the temperature of the photon, $T_\gamma$, and neutrino $T_{\nu}$ baths.  

As noted above, many axion models predict $Q$ decay into axions and ${\rm BR}_{a}$   in the $\rho_a$ Boltzmann equation (eq.~(\ref{eq:FBEqs_rhoa})) will no longer be zero. Given the huge mass difference between $Q$ and the axion, the resultant axion will be relativistic, unlike the axions produced through misalignment. Therefore, the energy density from eq.~\eqref{eq:FBEqs_rhoa} will contribute to the energy density of radiation today as dark radiation. An additional amount of relativistic degrees of freedom, $\DNeff$, is defined as,
\begin{equation}
\rho_\mathrm{R}\equiv \rho_\gamma\left[1+\frac{7}{8}\left(\frac{T_\nu}{T_\gamma}\right)(N_\mathrm{eff}^{\rm SM}+\Delta N_\mathrm{eff})\right]\,, 
\end{equation}
where $N_{\rm eff}^{\rm SM} = 3.045$ denotes the effective number of relativistic neutrinos in the SM \cite{deSalas:2016ztq}. Using the above equation together with $\rho_{\rm Rad} = \rho_{\rm R}^{\rm SM} + \rho_{a}$ allows us to solve for $\DNeff$
\begin{equation}
    \Delta N_{\rm eff} \equiv \left\{\frac{8}{7}\left(\frac{4}{11}\right)^{-\frac{4}{3}}+N_{\rm eff}^{\rm SM}\right\} 
 \frac{\rho_{a}}{\rho_{\rm R}^{\rm SM}}\,.
\end{equation}
Both $\rho_{\rm R}^{\rm SM}$ and $\rho_{a}$ are evaluated at the time of the CMB, around matter-radiation equality $T_{\rm eq}=0.75$ eV. We solve numerically the Friedmann-Boltzmann equations shown in Eq.~\eqref{eq:FBEqs} only until BBN ($T_{\rm BBN}$). The thermally decoupled axions will evolve independently of the SM and the energy densities of the two sectors can be easily extrapolated using entropy conservation

\begin{align}\label{eq:Neffg}
 \Delta N_{\rm eff} = \left\{ \frac{8}{7}\left(\frac{4}{11}\right)^{-\frac{4}{3}}+N_{\rm eff}^{\rm SM}\right\} 
 \frac{\rho_{a}(T_{\rm BBN})}{\rho_{\rm R}^{\rm SM}(T_{\rm BBN})}
 \left(\frac{g_*(T_{\rm BBN})}{g_*(T_{\rm eq})}\right)
 \left(\frac{g_{*S}(T_{\rm eq})}{g_{*S}(T_{\rm BBN})}\right)^{\frac{4}{3}}\,.
\end{align}\\

Additionally eq.~\eqref{eq:FBEqs_rhoa} contains the scattering terms between axions, the SM and heavy quarks. If these interactions are frequent enough, they will thermalise. Below the $m_Q$ threshold the thermal production rate is dominated by the dimension-5 anomalous term coupling the axion to the gluons~\cite{Graf:2010tv,Salvio:2013iaa}. Above $m_Q$, the scattering of the heavy quarks dominate even at temperatures much greater than this threshold as shown in Ref.~\cite{DEramo:2021lgb}. We have used the KSVZ axion production rate $\gamma_a$ from gluon scattering given in Ref.~\cite{DEramo:2021lgb} as well as the cross-sections in order to determine $\gamma_a$ for any $m_Q$ value. We do this primarily for ease and because KSVZ-type axion models are slightly more appealing because it is possible to achieve $N_{\rm DW}=1$ with them. When we make the association $m_Q\sim f_a$ we can find an approximate relation for when axion decoupling occurs, $\gamma_a\left(T_{\rm dec}^{a}\right)\approx 3 n_{a}^{\rm eq} H\left(T_{\rm dec}^{a}\right)$,
\be
T_{\rm dec}^{a}\approx 3.2\times 10^{-3}\, (m_Q)^{1.13} \,\,\,\,\,\, {\rm for} \,\,\,\, m_Q\sim f_a \gtrsim 10^{10} \, {\rm GeV}, 
\ee
where below $m_Q\sim 10^{10}\,{\rm GeV}$ the scaling is different, and is irrelevant for this work. Furthermore, considering the fact that, through EMD, the value for $f_a$ that results in the correct relic abundance is increased, we should keep in mind that the maximum value of \(f_a\) that can support thermalization at some point in the history is approximately given by \cite{Arias:2023wyg}
\begin{eqnarray}
    f_a \approx 3\times 10^{10}\,{\rm GeV} \sqrt{\frac{m_Q}{10^5\,{\rm GeV}}}.
\end{eqnarray}
With $m_Q\sim f_a$, axions will never equilibrate with the SM only when $f_a=m_Q\gtrsim10^{16}\,{\rm GeV}$, which is substantially higher than the value considered in this paper. Therefore we find it important to incorporate the axion production rate from the thermal plasma in our analysis.

\subsection{$\DNeff$ with standard cosmology ($d\leq 5$ models)}

By comparing the SM temperature at time of $Q$ decay $T_{\rm decay}^{Q}$, defined by $\Gamma_Q \approx H(T_{\rm decay}^Q)$, one can estimate whether the heavy quarks will decay prior to or after the axions have decoupled.\\ 

\noindent For $d=4$, the relation is approximately 
\be
\frac{T_{\rm decay}^{Q,\,d=4}}{T_{\rm dec}^{a}}\approx \left(\frac{1}{8\pi}\right)^{1/2}\left(\frac{1}{g_\star}\right)^{1/4}\left(\frac{90 M_{\rm Pl}^2}{8\pi^3}\right)^{1/4}\left(\frac{1}{3.2\times 10^{-3}\, (m_Q)^{1.13}}\right)\approx 16 \sqrt{\frac{M_{\rm Pl}}{m_Q}}
\ee
and since $M_{\rm Pl}> m_Q$, we have $T_{\rm decay}^{Q,\,d=4}>T_{\rm dec}^{a}$, meaning that we expect the standard $\DNeff\approx 0.027$, which itself is actually in reach of the CMB-HD projections. Furthermore, with the approximation $f_a\approx m_Q$, achieving the appropriate relic density for axion dark matter ($f_a\sim 10^{12}\,{\rm GeV}$) suggests $T_{\rm decay}^{Q,\,d=4}\sim 10^4 \,T_{\rm dec}^{a}$.

Since the thermal axion production rate $\gamma_a$ is determined by both $m_Q$ and $f_a$ independently, it is worth exploring the possibility where $m_Q\ll f_a$. This way one can envisage a situation where axions never couple through interactions mediated by the QCD anomaly term, such that $T_{\rm dec}^{a}>T_{\rm f}^Q$. However, in the dimension-4 case the decay rate is large enough to keep the axions in thermal equilibrium through decay and inverse decay of $Q$s as described by the following Boltzmann equation~\cite{Hall:2009bx,DEramo:2017ecx,Du:2021jcj}
\be
 \frac{\dd n_{a}}{\dd t} + 3H n_{a}  = \left(\frac{m^2 T}{2\pi^2}K_{1}\left(\frac{m_Q}{T}\right)\Gamma_{Q\rightarrow a + q} + \gamma_a\right) \left(1- \frac{n_a}{n_a^{\rm eq.}}\right)\,,\label{eq:FBEqs_d4}
\ee
where $K_1(m_Q/T)$ is the modified Bessel function of the second kind, and the $Q$ particles are produced in equilibrium at temperatures above and around freeze-out, $T_{\rm f}^Q\sim m_Q/25$. This has the effect of bringing the axions into equilibrium regardless of the Peccei-Quinn breaking scale. Therefore for $d=4$ preferred axion models the prediction of $\DNeff=0.027$ is fairly robust. Of course in works such as Ref.~\cite{DEramo:2021psx,DEramo:2021lgb,Arias:2023wyg}, it has been shown that in axion models where one does not explain dark matter one can enhance the number of relativistic degrees of freedom, but typically this is for values of $f_a$ substantially smaller than the values discussed in this work.\\

\noindent For dimension-5 models we can once again compare $T_{\rm decay}^{Q}$ and $T_{\rm dec}^{a}$, 
\be
\frac{T_{\rm decay}^{Q,\,d=5}}{T_{\rm dec}^{a}}\approx \left(\frac{m_Q^3}{16\pi\Lambda^2}\right)^{1/2}\left(\frac{1}{g_\star}\right)^{1/4}\left(\frac{90 M_{\rm Pl}^2}{8\pi^3}\right)^{1/4}\left(\frac{1}{3.2\times 10^{-3}\, (m_Q)^{1.13}}\right)\approx 10 \frac{\sqrt{m_QM_{\rm Pl}}}{\Lambda}
\ee
which suggests that, when taking the value $f_a\sim m_Q\sim 10^{12}\,{\rm GeV}$ that reproduces $\Omega_{\rm DM}$, $Q$ decay occurs after axions thermally decouple when $\Lambda\gtrsim 10^{16}\, {\rm GeV}$. Instead, if we set $\Lambda=M_{\rm Pl}$, we are met once again with a $m_Q/M_{\rm Pl}$ ratio, only this time it indicates that decays will occur after axion decoupling as long as $m_Q\lesssim M_{\rm Pl}/100$. Therefore, the decays of heavy $Q$s will directly alter the density of axions with respect to SM radiation, despite never dominating the Universe. 

\begin{figure}
    \centering
    \includegraphics[width=\textwidth]{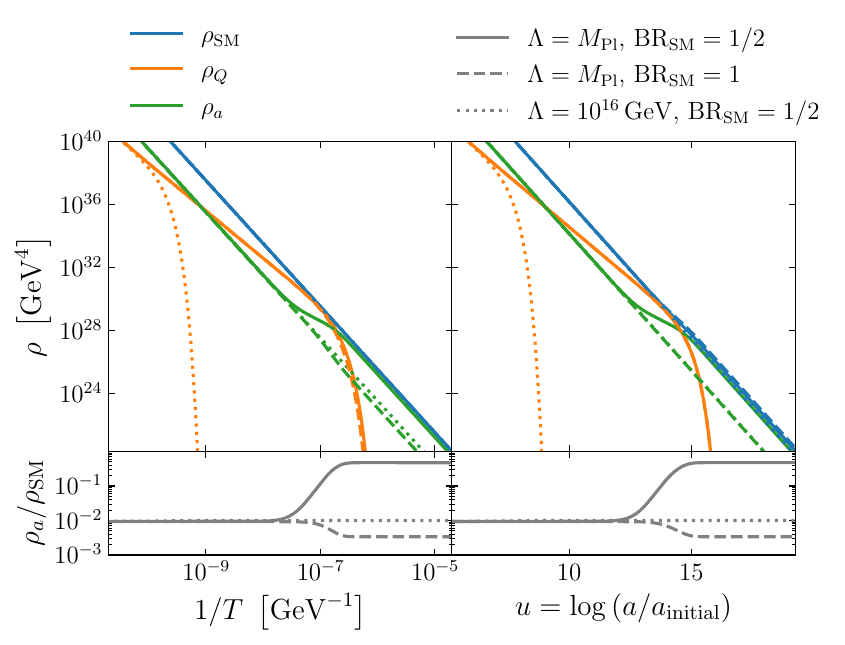}
    \caption{Top row shows the evolution of energy density for SM radiation (blue), heavy quarks, $Q$, (orange) and axions that contribute to dark radiation (green). We show three cases for $Q$ decays that proceed via dimension-5 non-renormalizable terms where the cutoff is $\Lambda=M_{\rm Pl}$ (solid and dashed lines for ${\rm BR}_{\rm SM}=1/2$ and ${\rm BR}_{\rm SM}=1$, respectively) or $\Lambda=10^{16}\,{\rm GeV}$ (dotted lines). We show how the branching ratio of $Q\rightarrow {\rm axions}$ affects the evolution. In the bottom row we show the ratio $\rho_{a}/\rho_{\rm SM}$ which contributes to $\DNeff$ Eq.~\eqref{eq:Neffg}. We take $m_Q\sim f_a$.  }
    \label{fig:evol_d5}
\end{figure}

In Fig.~\ref{fig:evol_d5} we show the evolution of the Universe for three $d=5$ cases, differentiated by the branching ratio of decay to the SM and by the decay cut-off scale, $\Lambda$, with all cases having $f_a= m_Q = 10^{12}\,{\rm GeV}$. The top row shows the SM, axion and heavy quark energy densities. On the left column the x-axis is the inverse temperature, on the right, it is the logarithm of the scale factor. Observe that with the maximum suppression $M_{\rm Pl}$ for decay, the $Q$s get quite close to dominating, as anticipated by our estimates in section \ref{sec:EQD}. Therefore the $Q$ decays will produce some meaningful impact on the SM radiation and dark radiation. We can see how the decay of $Q$ changes the SM radiation energy density by observing that on the top right panel the SM energy density in the case where ${\rm BR}_{\rm SM}=1$ is slightly higher than the ${\rm BR}_{\rm SM}=1/2$ case, because more of the $Q$ energy is being pumped into the SM bath. On the left this effect can be seen by noticing that the ${\rm BR}_{\rm SM}=1$ case has $\rho_Q$ depleting at a faster rate, but in reality it is just the bath $T$ decreasing more slowly. In the bottom row of Fig.~\ref{fig:evol_d5} the energy density ratios of $\rho_a/\rho_{\rm SM}$ are shown. The effect of $Q$ preferentially decaying into SM particles can be seen by observing that the ${\rm BR}_{\rm SM}=1$ case provides a substantially depleted ratio compared to the other two, which when plugged into eq.~(\ref{eq:Neffg}) gives $\DNeff=0.0097$, under half of the thermal, $d=4$ expectation. This situation is particularly relevant for many $d=5$ models because many charge configurations produce, to the lowest order, effective terms of the form
\be
\mathcal{L}_{Qf}^{d=5}=\overline{Q}_{\small L/R}\sigma_{\mu\nu}f_{\small R/L}G^{\mu\nu}
\ee
where $\sigma_{\mu\nu}=\frac{i}{2}\left[\gamma_\mu,\gamma_{\nu}\right]$ and $G^{\mu\nu}$ is the strong field strength tensor, and $f$ here is a SM fermion including leptons. These models predict ${\rm BR_{SM}}=1$ and therefore a $\DNeff<0.027$.  

The ${\rm BR}_{\rm SM}=1-{\rm BR}_{a}=1/2$ case produces an enhanced value of $\rho_a/\rho_{\rm SM}$ which corresponds to $\DNeff=1.37$. Current limits from PLANCK set $\DNeff\leq 0.244$ using the more stringent result (TT,TE,EE+low E)~\cite{Planck:2018vyg} which is much lower than the value predicted in our example, the corresponding limit on decays into axions is then ${\rm BR}_{a}\leq{0.12}$. With more sensitive probes expected such as the CMB-HD survey, substantial progress will be made. From Ref.~\cite{CMB-HD:2022bsz} we take the projection which brings the sensitivity to $\DNeff=0.014$. For our example with $m_Q=10^{12}\,{\rm GeV}$ and $\Lambda=M_{\rm Pl}$, these future limits will probe ${\rm BR}_{a}$ down to $\sim 0.01$, an order of magnitude improvement. As a reminder, we defined the `simplified' branching ratios, ${\rm BR}_{{\rm SM}/a}$, in a model-independent way, such that a value of ${\rm BR}_{a}=1/2$ corresponds to axion models with charge assignments such that there was only one decay $Q\rightarrow a + q$, i.e., half of the energy carried from a decay of $Q$ goes into the axion energy density. Because $Q$ is a fermion and the axion is a scalar ${\rm BR}_{a}$ should maximally be $1/2$.

\begin{figure}[t!]
    \centering
    \includegraphics[width=\textwidth]{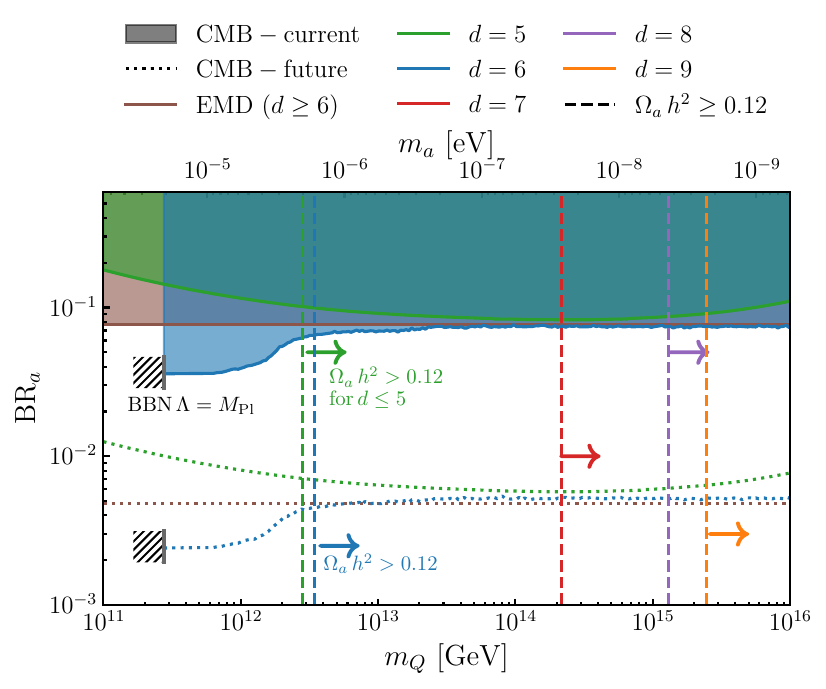}
    \caption{Existing limits and future projections on ${\rm BR}_{a}$ for different values of the dimension $d$ through which $Q$ decays proceed. Both limits and projections are taken from measurements of the CMB, current limits are shaded while future projections are denoted by dotted lines. We only show the specific limits for $d=5$ (green) and $d=6$ (blue) assuming $\Lambda=M_{\rm Pl}$. We show the general result when early matter domination (EMD) is achieved with brown lines because this is unchanged for $d\geq6$ models. We indicate where axions would be overproduced unless one introduces some fine-tuned $\theta_i$ for each dimension with a dashed vertical line and an arrow. }
    \label{fig:DNeff_d5}
\end{figure}

In Fig.~\ref{fig:DNeff_d5} we show how ${\rm BR}_{a}$ is constrained by current limits on $\DNeff$ provided by PLANCK~\cite{Planck:2018vyg} for different values of $m_Q$ and by assuming $\Lambda=M_{\rm Pl}$. We select $m_Q$ values in the region where $\Omega_a>\Omega_{\rm DM}$ to show how this limits the parameter space if we relax the restrictions on initial angle and allow for $\theta_i\ll 1$. It also shows the general behaviour of the limit. If a different $\Lambda$ was taken we expect the constraint to translate in the x-axis in the direction of lower $m_Q$ values. 

\subsection{$\DNeff$ with early matter domination (EMD) ($d\geq6$ Models)}

As shown in section~\ref{sec:EQD} preferred axion models that have $\mathcal{L}_{Qq}$ terms of the dimension $6$ and above can lead to a significant cosmological phase of EMD. As long as this happens, we can analytically approximate the implications $Q$ decay into axions will have for $\DNeff$. Furthermore, the extra powers of $\Lambda$ which suppress the decays of heavy quarks ensure that axions are not in thermal equilibrium by the time of decay. Assuming instantaneous decays, we can write the SM and axion energy density Boltzmann equations, 
\begin{subequations}\label{eq:FBEqs_EMD}
\bea
 \frac{\dd \rho_\mathrm{R}^{\rm SM}}{\dd t} + 4H \rho_\mathrm{R}^{\rm SM}&=&   {\rm BR}_{\rm SM}\Gamma_Q\rho_{Q}\,,\\
 \frac{\dd\rho_{a}}{\dd t} +  4H \rho_{a} &=& {\rm BR}_{a}\Gamma_Q\rho_{Q}\,.
\eea
\end{subequations}
as in \cref{eq:FBEqs} but we have dropped the axion annihilation term. Then by rewriting in terms of comoving energy density $a^4\rho$, we are left with
\be
\frac{\dd \left(a^4 \rho_\mathrm{R}^{\rm SM}\right)}{\dd t} = \frac{{\rm BR}_{\rm SM}}{{\rm BR}_{ a}}\frac{\dd \left(a^4 \rho_{a}\right)}{\dd t}, 
\ee
using the fact that both $\Gamma_{Q}$ and the branching ratios are not time-dependent. We integrate both sides of the equation, and use ${\rm BR}_{\rm SM}=1-{\rm BR}_{a}$ to get 
\be
\frac{\rho_\mathrm{R}^{\rm SM}}{\rho_a}=\frac{1-{\rm BR}_{a}}{{\rm BR}_{a}}\,.
\ee
With this relation we can use eq.~\eqref{eq:Neffg} to calculate $\DNeff$ at recombination. Note that in this approximation $\rho_{\rm R}^{\rm SM}/\rho_a$ is set at time of $Q$ decay, $T_{\rm decay}^Q$, not $T_{\rm BBN}$ as we do with our numerical calculation. The predicted $\DNeff$ is not very sensitive to the precise value of $T_{\rm decay}^Q$ because the main variation comes from the temperature dependence of $g_{\star}$ and $g_{\star S}$. There are three values of significance, $T_{\rm decay}^Q\sim 100\, {\rm GeV}$, for temperatures above which there is a minimal variation in the $g_\star$ values, $T_{\rm decay}^Q\sim T_{\rm QCD}$, below which bound state formation may alter the number density of $Q$, and $T_{\rm decay}^Q\sim T_{\rm BBN}$, below which constraints from BBN make the scenario unviable. Current limits from PLANCK set $\DNeff\leq 0.244$, which corresponds to an upper limit on the axion branching ratio ${\rm BR}_{a}\leq 0.077,\,0.057,\, 0.038$ for $T_{\rm decay}^Q\ = 100\,{\rm GeV}, \, T_{\rm QCD},\, T_{\rm BBN} $, respectively. For the CMB-HD survey, the projection of a sensitivity down to $\DNeff=0.014$~\cite{CMB-HD:2022bsz} translates to ${\rm BR}_{a}\leq 0.0048,\,0.0036,\, 0.0023,$ for the same respective temperatures. In Fig.~\ref{fig:DNeff_d5} we show this approximate result for $Q$ domination with $T_{\rm decay}^Q=100\,{\rm GeV}$ with a brown shaded region (currently excluded) and brown dotted line (projection).

Additionally, in Fig.~\ref{fig:DNeff_d5}  we plot the result for $d=6$ and $\Lambda=M_{\rm Pl}$ obtained by fully solving the system numerically in blue. We do this to compare our approximate result with the full calculation and see a remarkable level of agreement. The constraints from $d=6$ do start to drop below our approximation at around $\sim 5\times10^{12}\,{\rm GeV}$, but this is because $T_{\rm decay}^Q<100\,{\rm GeV}$ so there are fewer SM species in thermal equilibrium. 

For $d>6$ the results look similar to $d=6$, just the BBN constraint varies according to $d$ and $\Lambda$, so we only show the approximate result for EMD for these cases. As stressed above, $\Lambda$ is unknown and could be small enough for all dimensions considered to allow for $m_Q\sim10^{11}\,{\rm GeV}$, see Fig.~\ref{fig:scan_mQ_Lam}. Additionally, according to \cref{eq:est_MD} and Fig.~\ref{fig:diluca_similar} early $Q$ domination for this mass begins around $\sim 10^5\,{\rm GeV}$. So as long as $m_{Q}\gtrsim 10^6\,{\rm GeV}$ such that freeze-out occurs above the temperature of radiation and heavy quark equality, this limit may be applicable. Although with our assumption that $f_a\sim m_Q$, experimental constraints coming from supernovae~\cite{Brockway:1996yr, Grifols:1996id,Payez:2014xsa} and neutron star cooling~\cite{Hamaguchi:2018oqw,Leinson:2021ety} limit us to be above $m_Q\gtrsim10^9\,{\rm GeV}$. 

We also use our numerical solver to check what the predicted $\DNeff$ is for the case when ${\rm BR_{SM}}=1$. Here because the dominating quarks reheat the SM plasma and not the axions, the ratio $\rho_a/\rho_{\rm R}^{\rm SM}$ is going to be extremely suppressed. We find the result can be very low indeed, $\DNeff\approx 10^{-12}$, making such axions virtually undetectable. Naively one assumes that if the axion is ever in equilibrium is has at least $\DNeff\sim 0.027$, here we have an example where the axion was thermally coupled but diluted substantially by the decay of the heavy $Q$s. 

The vertical dashed lines in Fig.~\ref{fig:DNeff_d5} indicate where $\theta_i=0.5$ produces too many axions via misalignment $\Omega_a>\Omega_{\rm DM}$ for $d=6,7,8,9$ in colors blue, red, purple and orange, respectively. These values correspond to the points where the coloured shaded regions in Fig.~\ref{fig:scan_mQ_Lam} meet their relevant BBN constraint.

\section{Conclusions}
\label{sec:conclusion}
In this article we explore the role that heavy quarks required in many axion models may have on the evolution of the early Universe. In particular we find that these heavy quarks can dominate in the early Universe and dilute the relic cold axion abundance produced through misalignment. This has great consequence for our understanding of \textit{preferred axion models}, a subgroup of axion models that meet desirable criteria. Assuming standard early cosmology, the preferred axion models were believed to be limited to have heavy quark decay at most by dimension-5 effective operators. We have demonstrated that this is not the case and decays that proceed through effective operators up to dimension 9 still survive all known constraints.  

Our results here motivate the return to axion models where decays occur with effective operators of dimension $d\geq6$, since these should no longer be considered disfavored. In this article we have not attempted to collate these models in order to remain as general as possible. When we do study higher-dimensional axion models, we may find more models that meet with more restrictive criteria than considered here, such as no domain walls $N_{\rm DW}=1$. 

\begin{figure}
    \centering
    \includegraphics{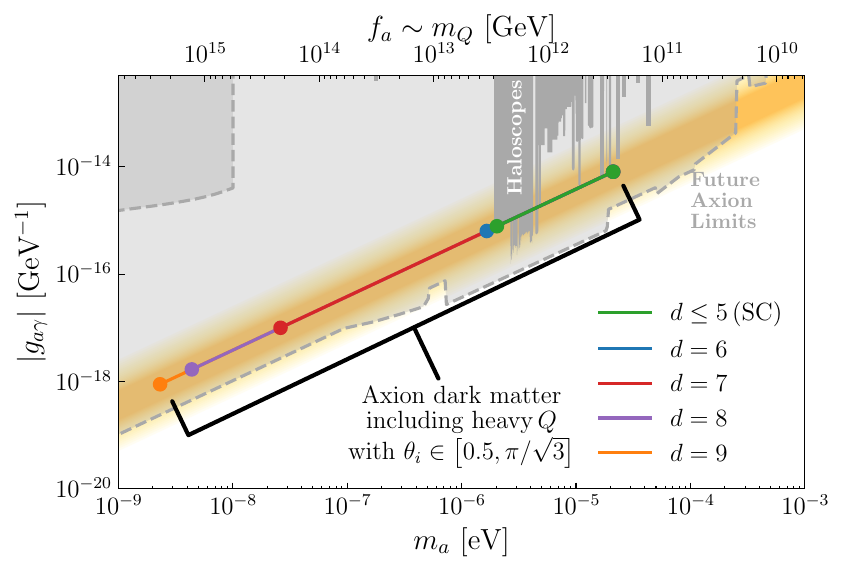}
    \caption{The extended region of preferred axion dark matter masses denoted by the colored lines, the previously understood window is shown in green, referring to decays via $d\leq 5$ effective operators and standard cosmology (SC). Decays that proceed through higher-dimensional effective operators and therefore produce a period of EMD are shown in blue ($d=6$), red ($d=7$), purple ($d=8$) and orange ($d=9$). We have remained agnostic about the suppression scale $\Lambda$. The maximum $m_a$ is the same for each case, and given by the rightmost green marker. The existing limits and future projections are taken from Ref.~\cite{ciaran_o_hare_2020_3932430}.}
    \label{fig:axion_limits}
\end{figure}

The effects of EMD on misalignment have been studied before, but the remarkable feature of this work is that we have not invoked an arbitrary matter field that dominates and decays before BBN, nothing additional needs to be added beyond what is already present in the axion model. Instead the heavy quarks required by many axion models actually produce this matter domination and it is therefore inaccurate to make conclusions about which models to discard without taking this into account. Importantly, this result has substantial consequences for axion dark matter and its expected mass range. In Fig.~\ref{fig:axion_limits} we show how the \textit{preferred axion dark matter} region is extended by this new insight, and it extends by three orders of magnitude on the low $m_a$ end. An exciting consequence of this result is that a large area of this parameter space is still viable but will be probed in the future by haloscope experiments. These experiments probe the axion-photon coupling $g_{a\gamma}$, thus we plot our results along the KSVZ line in the $\vert g_{a\gamma}\vert,$ vs. $m_a$ plane. Our results provide a consistent and well-motivated class of axion dark matter models to test. Furthermore, these mass values are achieved without fine-tuning of the initial misalignment angle and are entirely consistent with post-inflationary $U(1)_{\rm PQ}$ breaking. 

The axion limits and projections of Fig.~\ref{fig:axion_limits} have been plotted using Ref.~\cite{ciaran_o_hare_2020_3932430} and are indicated by dark and light gray shaded regions, respectively. The limits come from microwave cavity experiments ADMX~\cite{ADMX:2009iij,ADMX:2019uok} and HAYSTAC~\cite{HAYSTAC:2018rwy}, whereas the future projections comprise of results from a broad group of operational or proposed experiments~\cite{Irastorza:2018dyq, Sikivie:2020zpn, Billard:2021uyg, Adams:2022pbo}, some of which are based on the microwave cavity technique, i.e. CULTASK~\cite{Lee:2020cfj,CAPP:2020utb}, KLASH~\cite{Alesini:2017ifp}, ORGAN~\cite{McAllister:2017lkb}, RADES~\cite{Melcon:2018dba}, and QUAX~\cite{Alesini:2020vny}. Experiments using a different technology but probing a similar mass range are MADMAX~\cite{MADMAX:2019pub}, BABYIAXO~\cite{Ahyoune:2023gfw} and FLASH~\cite{Alesini:2017ifp}. For lower axion masses in the $10^{-9}$ to $10^{-7}$ eV range, experiments such as SRF~\cite{Berlin:2020vrk, Tang:2023oid}, DMRadio~\cite{DMRadio:2022pkf}, and ABRACADABRA~\cite{Salemi:2021gck} show particular promise.  

Since the matter-dominating field in this scenario may actually decay into axions, the phase of early $Q$ domination will alter the expectation for the axion contribution to the number of relativistic degrees of freedom. This actually gives us another experimental window into preferred axion models with $d\geq5$. Probing the relative decays into axions and SM particles may allow further discrimination of models. In particular, all $d=4$ models are expected to reproduce the equilibrium result $\DNeff\approx0.027$, but many $d=5$ predict a ${\rm BR_{SM}}=1$ and predict a value less than half of the equilibrium result, $\DNeff\approx0.0097$. As mentioned above, further study on $d\geq 6$ is required but models may indicate a specific value of ${\rm BR_{SM}}$ which could already be constrained by experiment or be in reach of future probes. In the event of discovering axion dark matter in the non-standard mass window, this information would be pivotal for determining whether $U(1)_{\rm PQ}$ breaking occurred before or after inflation.

\begin{acknowledgments}
AC would like to thank Dorian Amaral, Ken Mimasu, Matthew Kirk, Luca Di Luzio, and Bradley Kavanagh for useful discussions. JO would like to thank Paola Arias and Moira Venegas for discussions regarding thermally produced axions. This work is supported by the grant ``AstroCeNT: Particle Astrophysics Science and Technology Centre" carried out within the International Research Agendas programme of the Foundation for Polish Science financed by the European Union under the European Regional Development Fund. 
\end{acknowledgments}

\bibliographystyle{JHEP}
\bibliography{ref}

\providecommand{\href}[2]{#2}\begingroup\raggedright\begin{thebibliography}{100}

\bibitem{Peccei:1977hh}
R.D.~Peccei and H.R.~Quinn, \emph{{CP Conservation in the Presence of
  Instantons}}, \href{https://doi.org/10.1103/PhysRevLett.38.1440}{\emph{Phys.
  Rev. Lett.} {\bfseries 38} (1977) 1440}.

\bibitem{Peccei:1977np}
R.D.~Peccei and H.R.~Quinn, \emph{{Some Aspects of Instantons}},
  \href{https://doi.org/10.1007/BF02730110}{\emph{Nuovo Cim. A} {\bfseries 41}
  (1977) 309}.

\bibitem{Preskill:1982cy}
J.~Preskill, M.B.~Wise and F.~Wilczek, \emph{{Cosmology of the Invisible
  Axion}}, \href{https://doi.org/10.1016/0370-2693(83)90637-8}{\emph{Phys.
  Lett. B} {\bfseries 120} (1983) 127}.

\bibitem{Dine:1982ah}
M.~Dine and W.~Fischler, \emph{{The Not So Harmless Axion}},
  \href{https://doi.org/10.1016/0370-2693(83)90639-1}{\emph{Phys. Lett. B}
  {\bfseries 120} (1983) 137}.

\bibitem{Abbott:1982af}
L.F.~Abbott and P.~Sikivie, \emph{{A Cosmological Bound on the Invisible
  Axion}}, \href{https://doi.org/10.1016/0370-2693(83)90638-X}{\emph{Phys.
  Lett. B} {\bfseries 120} (1983) 133}.

\bibitem{Raffelt:1990yz}
G.G.~Raffelt, \emph{{Astrophysical methods to constrain axions and other novel
  particle phenomena}},
  \href{https://doi.org/10.1016/0370-1573(90)90054-6}{\emph{Phys. Rept.}
  {\bfseries 198} (1990) 1}.

\bibitem{Raffelt:2006cw}
G.G.~Raffelt, \emph{{Astrophysical axion bounds}},
  \href{https://doi.org/10.1007/978-3-540-73518-2_3}{\emph{Lect. Notes Phys.}
  {\bfseries 741} (2008) 51}
  [\href{https://arxiv.org/abs/hep-ph/0611350}{{\ttfamily hep-ph/0611350}}].

\bibitem{DiLuzio:2020wdo}
L.~Di~Luzio, M.~Giannotti, E.~Nardi and L.~Visinelli, \emph{{The landscape of
  QCD axion models}},
  \href{https://doi.org/10.1016/j.physrep.2020.06.002}{\emph{Phys. Rept.}
  {\bfseries 870} (2020) 1} [\href{https://arxiv.org/abs/2003.01100}{{\ttfamily
  2003.01100}}].

\bibitem{Sikivie:1983ip}
P.~Sikivie, \emph{{Experimental Tests of the Invisible Axion}},
  \href{https://doi.org/10.1103/PhysRevLett.51.1415}{\emph{Phys. Rev. Lett.}
  {\bfseries 51} (1983) 1415}.

\bibitem{ADMX:2009iij}
{\scshape ADMX} collaboration, \emph{{A SQUID-based microwave cavity search for
  dark-matter axions}},
  \href{https://doi.org/10.1103/PhysRevLett.104.041301}{\emph{Phys. Rev. Lett.}
  {\bfseries 104} (2010) 041301}
  [\href{https://arxiv.org/abs/0910.5914}{{\ttfamily 0910.5914}}].

\bibitem{ADMX:2019uok}
{\scshape ADMX} collaboration, \emph{{Extended Search for the Invisible Axion
  with the Axion Dark Matter Experiment}},
  \href{https://doi.org/10.1103/PhysRevLett.124.101303}{\emph{Phys. Rev. Lett.}
  {\bfseries 124} (2020) 101303}
  [\href{https://arxiv.org/abs/1910.08638}{{\ttfamily 1910.08638}}].

\bibitem{HAYSTAC:2018rwy}
{\scshape HAYSTAC} collaboration, \emph{{Results from phase 1 of the HAYSTAC
  microwave cavity axion experiment}},
  \href{https://doi.org/10.1103/PhysRevD.97.092001}{\emph{Phys. Rev. D}
  {\bfseries 97} (2018) 092001}
  [\href{https://arxiv.org/abs/1803.03690}{{\ttfamily 1803.03690}}].

\bibitem{Irastorza:2018dyq}
I.G.~Irastorza and J.~Redondo, \emph{{New experimental approaches in the search
  for axion-like particles}},
  \href{https://doi.org/10.1016/j.ppnp.2018.05.003}{\emph{Prog. Part. Nucl.
  Phys.} {\bfseries 102} (2018) 89}
  [\href{https://arxiv.org/abs/1801.08127}{{\ttfamily 1801.08127}}].

\bibitem{Sikivie:2020zpn}
P.~Sikivie, \emph{{Invisible Axion Search Methods}},
  \href{https://doi.org/10.1103/RevModPhys.93.015004}{\emph{Rev. Mod. Phys.}
  {\bfseries 93} (2021) 015004}
  [\href{https://arxiv.org/abs/2003.02206}{{\ttfamily 2003.02206}}].

\bibitem{Billard:2021uyg}
J.~Billard et~al., \emph{{Direct detection of dark matter\textemdash{}APPEC
  committee report}},
  \href{https://doi.org/10.1088/1361-6633/ac5754}{\emph{Rept. Prog. Phys.}
  {\bfseries 85} (2022) 056201}
  [\href{https://arxiv.org/abs/2104.07634}{{\ttfamily 2104.07634}}].

\bibitem{Adams:2022pbo}
C.B.~Adams et~al., \emph{{Axion Dark Matter}},  in \emph{{Snowmass 2021}}, 3,
  2022 [\href{https://arxiv.org/abs/2203.14923}{{\ttfamily 2203.14923}}].

\bibitem{DiLuzio:2016sbl}
L.~Di~Luzio, F.~Mescia and E.~Nardi, \emph{{Redefining the Axion Window}},
  \href{https://doi.org/10.1103/PhysRevLett.118.031801}{\emph{Phys. Rev. Lett.}
  {\bfseries 118} (2017) 031801}
  [\href{https://arxiv.org/abs/1610.07593}{{\ttfamily 1610.07593}}].

\bibitem{DiLuzio:2017pfr}
L.~Di~Luzio, F.~Mescia and E.~Nardi, \emph{{Window for preferred axion
  models}}, \href{https://doi.org/10.1103/PhysRevD.96.075003}{\emph{Phys. Rev.
  D} {\bfseries 96} (2017) 075003}
  [\href{https://arxiv.org/abs/1705.05370}{{\ttfamily 1705.05370}}].

\bibitem{Steinhardt:1983ia}
P.J.~Steinhardt and M.S.~Turner, \emph{{Saving the Invisible Axion}},
  \href{https://doi.org/10.1016/0370-2693(83)90727-X}{\emph{Phys. Lett. B}
  {\bfseries 129} (1983) 51}.

\bibitem{Lazarides:1990xp}
G.~Lazarides, R.K.~Schaefer, D.~Seckel and Q.~Shafi, \emph{{Dilution of
  Cosmological Axions by Entropy Production}},
  \href{https://doi.org/10.1016/0550-3213(90)90244-8}{\emph{Nucl. Phys. B}
  {\bfseries 346} (1990) 193}.

\bibitem{Kawasaki:1995vt}
M.~Kawasaki, T.~Moroi and T.~Yanagida, \emph{{Can decaying particles raise the
  upper bound on the Peccei-Quinn scale?}},
  \href{https://doi.org/10.1016/0370-2693(96)00743-5}{\emph{Phys. Lett. B}
  {\bfseries 383} (1996) 313}
  [\href{https://arxiv.org/abs/hep-ph/9510461}{{\ttfamily hep-ph/9510461}}].

\bibitem{Ramberg:2019dgi}
N.~Ramberg and L.~Visinelli, \emph{{Probing the Early Universe with Axion
  Physics and Gravitational Waves}},
  \href{https://doi.org/10.1103/PhysRevD.99.123513}{\emph{Phys. Rev. D}
  {\bfseries 99} (2019) 123513}
  [\href{https://arxiv.org/abs/1904.05707}{{\ttfamily 1904.05707}}].

\bibitem{Arias:2021rer}
P.~Arias, N.~Bernal, D.~Karamitros, C.~Maldonado, L.~Roszkowski and M.~Venegas,
  \emph{{New opportunities for axion dark matter searches in nonstandard
  cosmological models}},
  \href{https://doi.org/10.1088/1475-7516/2021/11/003}{\emph{JCAP} {\bfseries
  11} (2021) 003} [\href{https://arxiv.org/abs/2107.13588}{{\ttfamily
  2107.13588}}].

\bibitem{Arias:2022qjt}
P.~Arias, N.~Bernal, J.K.~Osi\'nski and L.~Roszkowski, \emph{{Dark matter
  axions in the early universe with a period of increasing temperature}},
  \href{https://doi.org/10.1088/1475-7516/2023/05/028}{\emph{JCAP} {\bfseries
  05} (2023) 028} [\href{https://arxiv.org/abs/2207.07677}{{\ttfamily
  2207.07677}}].

\bibitem{Kim:2008hd}
J.E.~Kim and G.~Carosi, \emph{{Axions and the Strong CP Problem}},
  \href{https://doi.org/10.1103/RevModPhys.82.557}{\emph{Rev. Mod. Phys.}
  {\bfseries 82} (2010) 557} [\href{https://arxiv.org/abs/0807.3125}{{\ttfamily
  0807.3125}}].

\bibitem{Kim:1979if}
J.E.~Kim, \emph{{Weak Interaction Singlet and Strong CP Invariance}},
  \href{https://doi.org/10.1103/PhysRevLett.43.103}{\emph{Phys. Rev. Lett.}
  {\bfseries 43} (1979) 103}.

\bibitem{Shifman:1979if}
M.A.~Shifman, A.I.~Vainshtein and V.I.~Zakharov, \emph{{Can Confinement Ensure
  Natural CP Invariance of Strong Interactions?}},
  \href{https://doi.org/10.1016/0550-3213(80)90209-6}{\emph{Nucl. Phys. B}
  {\bfseries 166} (1980) 493}.

\bibitem{Zhitnitsky:1980tq}
A.R.~Zhitnitsky, \emph{{On Possible Suppression of the Axion Hadron
  Interactions. (In Russian)}}, {\emph{Sov. J. Nucl. Phys.} {\bfseries 31}
  (1980) 260}.

\bibitem{Dine:1981rt}
M.~Dine, W.~Fischler and M.~Srednicki, \emph{{A Simple Solution to the Strong
  CP Problem with a Harmless Axion}},
  \href{https://doi.org/10.1016/0370-2693(81)90590-6}{\emph{Phys. Lett. B}
  {\bfseries 104} (1981) 199}.

\bibitem{Kawasaki:2004qu}
M.~Kawasaki, K.~Kohri and T.~Moroi, \emph{{Big-Bang nucleosynthesis and
  hadronic decay of long-lived massive particles}},
  \href{https://doi.org/10.1103/PhysRevD.71.083502}{\emph{Phys. Rev. D}
  {\bfseries 71} (2005) 083502}
  [\href{https://arxiv.org/abs/astro-ph/0408426}{{\ttfamily
  astro-ph/0408426}}].

\bibitem{Jedamzik:2006xz}
K.~Jedamzik, \emph{{Big bang nucleosynthesis constraints on hadronically and
  electromagnetically decaying relic neutral particles}},
  \href{https://doi.org/10.1103/PhysRevD.74.103509}{\emph{Phys. Rev. D}
  {\bfseries 74} (2006) 103509}
  [\href{https://arxiv.org/abs/hep-ph/0604251}{{\ttfamily hep-ph/0604251}}].

\bibitem{Jedamzik:2007qk}
K.~Jedamzik, \emph{{Bounds on long-lived charged massive particles from Big
  Bang nucleosynthesis}},
  \href{https://doi.org/10.1088/1475-7516/2008/03/008}{\emph{JCAP} {\bfseries
  03} (2008) 008} [\href{https://arxiv.org/abs/0710.5153}{{\ttfamily
  0710.5153}}].

\bibitem{Kawasaki:2017bqm}
M.~Kawasaki, K.~Kohri, T.~Moroi and Y.~Takaesu, \emph{{Revisiting Big-Bang
  Nucleosynthesis Constraints on Long-Lived Decaying Particles}},
  \href{https://doi.org/10.1103/PhysRevD.97.023502}{\emph{Phys. Rev. D}
  {\bfseries 97} (2018) 023502}
  [\href{https://arxiv.org/abs/1709.01211}{{\ttfamily 1709.01211}}].

\bibitem{DeLuca:2018mzn}
V.~De~Luca, A.~Mitridate, M.~Redi, J.~Smirnov and A.~Strumia, \emph{{Colored
  Dark Matter}}, \href{https://doi.org/10.1103/PhysRevD.97.115024}{\emph{Phys.
  Rev. D} {\bfseries 97} (2018) 115024}
  [\href{https://arxiv.org/abs/1801.01135}{{\ttfamily 1801.01135}}].

\bibitem{Gross:2018zha}
C.~Gross, A.~Mitridate, M.~Redi, J.~Smirnov and A.~Strumia, \emph{{Cosmological
  Abundance of Colored Relics}},
  \href{https://doi.org/10.1103/PhysRevD.99.016024}{\emph{Phys. Rev. D}
  {\bfseries 99} (2019) 016024}
  [\href{https://arxiv.org/abs/1811.08418}{{\ttfamily 1811.08418}}].

\bibitem{Dover:1979sn}
C.B.~Dover, T.K.~Gaisser and G.~Steigman, \emph{{COSMOLOGICAL CONSTRAINTS ON
  NEW STABLE HADRONS}},
  \href{https://doi.org/10.1103/PhysRevLett.42.1117}{\emph{Phys. Rev. Lett.}
  {\bfseries 42} (1979) 1117}.

\bibitem{Griest:1989wd}
K.~Griest and M.~Kamionkowski, \emph{{Unitarity Limits on the Mass and Radius
  of Dark Matter Particles}},
  \href{https://doi.org/10.1103/PhysRevLett.64.615}{\emph{Phys. Rev. Lett.}
  {\bfseries 64} (1990) 615}.

\bibitem{Nardi:1990ku}
E.~Nardi and E.~Roulet, \emph{{Are exotic stable quarks cosmologically
  allowed?}}, \href{https://doi.org/10.1016/0370-2693(90)90172-3}{\emph{Phys.
  Lett. B} {\bfseries 245} (1990) 105}.

\bibitem{Arvanitaki:2005fa}
A.~Arvanitaki, C.~Davis, P.W.~Graham, A.~Pierce and J.G.~Wacker, \emph{{Limits
  on split supersymmetry from gluino cosmology}},
  \href{https://doi.org/10.1103/PhysRevD.72.075011}{\emph{Phys. Rev. D}
  {\bfseries 72} (2005) 075011}
  [\href{https://arxiv.org/abs/hep-ph/0504210}{{\ttfamily hep-ph/0504210}}].

\bibitem{Kang:2006yd}
J.~Kang, M.A.~Luty and S.~Nasri, \emph{{The Relic abundance of long-lived heavy
  colored particles}},
  \href{https://doi.org/10.1088/1126-6708/2008/09/086}{\emph{JHEP} {\bfseries
  09} (2008) 086} [\href{https://arxiv.org/abs/hep-ph/0611322}{{\ttfamily
  hep-ph/0611322}}].

\bibitem{Jacoby:2007nw}
C.~Jacoby and S.~Nussinov, \emph{{The Relic Abundance of Massive Colored
  Particles after a Late Hadronic Annihilation Stage}},
  \href{https://arxiv.org/abs/0712.2681}{{\ttfamily 0712.2681}}.

\bibitem{Kusakabe:2011hk}
M.~Kusakabe and T.~Takesako, \emph{{Resonant annihilation of long-lived massive
  colored particles through hadronic collisions}},
  \href{https://doi.org/10.1103/PhysRevD.85.015005}{\emph{Phys. Rev. D}
  {\bfseries 85} (2012) 015005}
  [\href{https://arxiv.org/abs/1112.0860}{{\ttfamily 1112.0860}}].

\bibitem{Perl:2001xi}
M.L.~Perl, P.C.~Kim, V.~Halyo, E.R.~Lee, I.T.~Lee, D.~Loomba et~al., \emph{{The
  Search for stable, massive, elementary particles}},
  \href{https://doi.org/10.1142/S0217751X01003548}{\emph{Int. J. Mod. Phys. A}
  {\bfseries 16} (2001) 2137}
  [\href{https://arxiv.org/abs/hep-ex/0102033}{{\ttfamily hep-ex/0102033}}].

\bibitem{Hertzberg:2016jie}
M.P.~Hertzberg and A.~Masoumi, \emph{{Astrophysical Constraints on Singlet
  Scalars at LHC}},
  \href{https://doi.org/10.1088/1475-7516/2017/04/028}{\emph{JCAP} {\bfseries
  04} (2017) 028} [\href{https://arxiv.org/abs/1607.06445}{{\ttfamily
  1607.06445}}].

\bibitem{Gould:1989gw}
A.~Gould, B.T.~Draine, R.W.~Romani and S.~Nussinov, \emph{{Neutron Stars:
  Graveyard of Charged Dark Matter}},
  \href{https://doi.org/10.1016/0370-2693(90)91745-W}{\emph{Phys. Lett. B}
  {\bfseries 238} (1990) 337}.

\bibitem{Mack:2007xj}
G.D.~Mack, J.F.~Beacom and G.~Bertone, \emph{{Towards Closing the Window on
  Strongly Interacting Dark Matter: Far-Reaching Constraints from Earth's Heat
  Flow}}, \href{https://doi.org/10.1103/PhysRevD.76.043523}{\emph{Phys. Rev. D}
  {\bfseries 76} (2007) 043523}
  [\href{https://arxiv.org/abs/0705.4298}{{\ttfamily 0705.4298}}].

\bibitem{Allahverdi:2020bys}
R.~Allahverdi et~al., \emph{{The First Three Seconds: a Review of Possible
  Expansion Histories of the Early Universe}},
  \href{https://arxiv.org/abs/2006.16182}{{\ttfamily 2006.16182}}.

\bibitem{Kawasaki:2000en}
M.~Kawasaki, K.~Kohri and N.~Sugiyama, \emph{{MeV scale reheating temperature
  and thermalization of neutrino background}},
  \href{https://doi.org/10.1103/PhysRevD.62.023506}{\emph{Phys. Rev. D}
  {\bfseries 62} (2000) 023506}
  [\href{https://arxiv.org/abs/astro-ph/0002127}{{\ttfamily
  astro-ph/0002127}}].

\bibitem{Hannestad:2004px}
S.~Hannestad, \emph{{What is the lowest possible reheating temperature?}},
  \href{https://doi.org/10.1103/PhysRevD.70.043506}{\emph{Phys. Rev. D}
  {\bfseries 70} (2004) 043506}
  [\href{https://arxiv.org/abs/astro-ph/0403291}{{\ttfamily
  astro-ph/0403291}}].

\bibitem{Ichikawa:2005vw}
K.~Ichikawa, M.~Kawasaki and F.~Takahashi, \emph{{The Oscillation effects on
  thermalization of the neutrinos in the Universe with low reheating
  temperature}}, \href{https://doi.org/10.1103/PhysRevD.72.043522}{\emph{Phys.
  Rev. D} {\bfseries 72} (2005) 043522}
  [\href{https://arxiv.org/abs/astro-ph/0505395}{{\ttfamily
  astro-ph/0505395}}].

\bibitem{Ichikawa:2006vm}
K.~Ichikawa, M.~Kawasaki and F.~Takahashi, \emph{{Constraint on the Effective
  Number of Neutrino Species from the WMAP and SDSS LRG Power Spectra}},
  \href{https://doi.org/10.1088/1475-7516/2007/05/007}{\emph{JCAP} {\bfseries
  05} (2007) 007} [\href{https://arxiv.org/abs/astro-ph/0611784}{{\ttfamily
  astro-ph/0611784}}].

\bibitem{deSalas:2015glj}
P.F.~de~Salas, M.~Lattanzi, G.~Mangano, G.~Miele, S.~Pastor and O.~Pisanti,
  \emph{{Bounds on very low reheating scenarios after Planck}},
  \href{https://doi.org/10.1103/PhysRevD.92.123534}{\emph{Phys. Rev. D}
  {\bfseries 92} (2015) 123534}
  [\href{https://arxiv.org/abs/1511.00672}{{\ttfamily 1511.00672}}].

\bibitem{Hasegawa:2019jsa}
T.~Hasegawa, N.~Hiroshima, K.~Kohri, R.S.L.~Hansen, T.~Tram and S.~Hannestad,
  \emph{{MeV-scale reheating temperature and thermalization of oscillating
  neutrinos by radiative and hadronic decays of massive particles}},
  \href{https://doi.org/10.1088/1475-7516/2019/12/012}{\emph{JCAP} {\bfseries
  12} (2019) 012} [\href{https://arxiv.org/abs/1908.10189}{{\ttfamily
  1908.10189}}].

\bibitem{Sikivie:2006ni}
P.~Sikivie, \emph{{Axion Cosmology}},
  \href{https://doi.org/10.1007/978-3-540-73518-2_2}{\emph{Lect. Notes Phys.}
  {\bfseries 741} (2008) 19}
  [\href{https://arxiv.org/abs/astro-ph/0610440}{{\ttfamily
  astro-ph/0610440}}].

\bibitem{Marsh:2015xka}
D.J.E.~Marsh, \emph{{Axion Cosmology}},
  \href{https://doi.org/10.1016/j.physrep.2016.06.005}{\emph{Phys. Rept.}
  {\bfseries 643} (2016) 1} [\href{https://arxiv.org/abs/1510.07633}{{\ttfamily
  1510.07633}}].

\bibitem{Planck:2018vyg}
{\scshape Planck} collaboration, \emph{{Planck 2018 results. VI. Cosmological
  parameters}},
  \href{https://doi.org/10.1051/0004-6361/201833910}{\emph{Astron. Astrophys.}
  {\bfseries 641} (2020) A6}
  [\href{https://arxiv.org/abs/1807.06209}{{\ttfamily 1807.06209}}].

\bibitem{Kamionkowski:1992mf}
M.~Kamionkowski and J.~March-Russell, \emph{{Planck scale physics and the
  Peccei-Quinn mechanism}},
  \href{https://doi.org/10.1016/0370-2693(92)90492-M}{\emph{Phys. Lett. B}
  {\bfseries 282} (1992) 137}
  [\href{https://arxiv.org/abs/hep-th/9202003}{{\ttfamily hep-th/9202003}}].

\bibitem{Holman:1992us}
R.~Holman, S.D.H.~Hsu, T.W.~Kephart, E.W.~Kolb, R.~Watkins and L.M.~Widrow,
  \emph{{Solutions to the strong CP problem in a world with gravity}},
  \href{https://doi.org/10.1016/0370-2693(92)90491-L}{\emph{Phys. Lett. B}
  {\bfseries 282} (1992) 132}
  [\href{https://arxiv.org/abs/hep-ph/9203206}{{\ttfamily hep-ph/9203206}}].

\bibitem{Barr:1992qq}
S.M.~Barr and D.~Seckel, \emph{{Planck scale corrections to axion models}},
  \href{https://doi.org/10.1103/PhysRevD.46.539}{\emph{Phys. Rev. D} {\bfseries
  46} (1992) 539}.

\bibitem{Ringwald:2015dsf}
A.~Ringwald and K.~Saikawa, \emph{{Axion dark matter in the post-inflationary
  Peccei-Quinn symmetry breaking scenario}},
  \href{https://doi.org/10.1103/PhysRevD.93.085031}{\emph{Phys. Rev. D}
  {\bfseries 93} (2016) 085031}
  [\href{https://arxiv.org/abs/1512.06436}{{\ttfamily 1512.06436}}].

\bibitem{Hertzberg:2008wr}
M.P.~Hertzberg, M.~Tegmark and F.~Wilczek, \emph{{Axion Cosmology and the
  Energy Scale of Inflation}},
  \href{https://doi.org/10.1103/PhysRevD.78.083507}{\emph{Phys. Rev. D}
  {\bfseries 78} (2008) 083507}
  [\href{https://arxiv.org/abs/0807.1726}{{\ttfamily 0807.1726}}].

\bibitem{Kibble:1976sj}
T.W.B.~Kibble, \emph{{Topology of Cosmic Domains and Strings}},
  \href{https://doi.org/10.1088/0305-4470/9/8/029}{\emph{J. Phys. A} {\bfseries
  9} (1976) 1387}.

\bibitem{Kibble:1982dd}
T.W.B.~Kibble, G.~Lazarides and Q.~Shafi, \emph{{Walls Bounded by Strings}},
  \href{https://doi.org/10.1103/PhysRevD.26.435}{\emph{Phys. Rev. D} {\bfseries
  26} (1982) 435}.

\bibitem{Vilenkin:2000jqa}
A.~Vilenkin and E.P.S.~Shellard, \emph{{Cosmic Strings and Other Topological
  Defects}}, Cambridge University Press (7, 2000).

\bibitem{Hagmann:2000ja}
C.~Hagmann, S.~Chang and P.~Sikivie, \emph{{Axion radiation from strings}},
  \href{https://doi.org/10.1103/PhysRevD.63.125018}{\emph{Phys. Rev. D}
  {\bfseries 63} (2001) 125018}
  [\href{https://arxiv.org/abs/hep-ph/0012361}{{\ttfamily hep-ph/0012361}}].

\bibitem{Wantz:2009it}
O.~Wantz and E.P.S.~Shellard, \emph{{Axion Cosmology Revisited}},
  \href{https://doi.org/10.1103/PhysRevD.82.123508}{\emph{Phys. Rev. D}
  {\bfseries 82} (2010) 123508}
  [\href{https://arxiv.org/abs/0910.1066}{{\ttfamily 0910.1066}}].

\bibitem{Hiramatsu:2010yu}
T.~Hiramatsu, M.~Kawasaki, T.~Sekiguchi, M.~Yamaguchi and J.~Yokoyama,
  \emph{{Improved estimation of radiated axions from cosmological axionic
  strings}}, \href{https://doi.org/10.1103/PhysRevD.83.123531}{\emph{Phys. Rev.
  D} {\bfseries 83} (2011) 123531}
  [\href{https://arxiv.org/abs/1012.5502}{{\ttfamily 1012.5502}}].

\bibitem{Kawasaki:2014sqa}
M.~Kawasaki, K.~Saikawa and T.~Sekiguchi, \emph{{Axion dark matter from
  topological defects}},
  \href{https://doi.org/10.1103/PhysRevD.91.065014}{\emph{Phys. Rev. D}
  {\bfseries 91} (2015) 065014}
  [\href{https://arxiv.org/abs/1412.0789}{{\ttfamily 1412.0789}}].

\bibitem{Gorghetto:2018ocs}
M.~Gorghetto and G.~Villadoro, \emph{{Topological Susceptibility and QCD Axion
  Mass: QED and NNLO corrections}},
  \href{https://doi.org/10.1007/JHEP03(2019)033}{\emph{JHEP} {\bfseries 03}
  (2019) 033} [\href{https://arxiv.org/abs/1812.01008}{{\ttfamily
  1812.01008}}].

\bibitem{Buschmann:2021sdq}
M.~Buschmann, J.W.~Foster, A.~Hook, A.~Peterson, D.E.~Willcox, W.~Zhang et~al.,
  \emph{{Dark matter from axion strings with adaptive mesh refinement}},
  \href{https://doi.org/10.1038/s41467-022-28669-y}{\emph{Nature Commun.}
  {\bfseries 13} (2022) 1049}
  [\href{https://arxiv.org/abs/2108.05368}{{\ttfamily 2108.05368}}].

\bibitem{Sikivie:1982qv}
P.~Sikivie, \emph{{Of Axions, Domain Walls and the Early Universe}},
  \href{https://doi.org/10.1103/PhysRevLett.48.1156}{\emph{Phys. Rev. Lett.}
  {\bfseries 48} (1982) 1156}.

\bibitem{Vilenkin:1982ks}
A.~Vilenkin and A.E.~Everett, \emph{{Cosmic Strings and Domain Walls in Models
  with Goldstone and PseudoGoldstone Bosons}},
  \href{https://doi.org/10.1103/PhysRevLett.48.1867}{\emph{Phys. Rev. Lett.}
  {\bfseries 48} (1982) 1867}.

\bibitem{Barr:1986hs}
S.M.~Barr, K.~Choi and J.E.~Kim, \emph{{Some aspects of axion cosmology in
  unified and superstring models}},
  \href{https://doi.org/10.1016/0550-3213(87)90288-4}{\emph{Nucl. Phys. B}
  {\bfseries 283} (1987) 591}.

\bibitem{Turner:1985si}
M.S.~Turner, \emph{{Cosmic and Local Mass Density of Invisible Axions}},
  \href{https://doi.org/10.1103/PhysRevD.33.889}{\emph{Phys. Rev. D} {\bfseries
  33} (1986) 889}.

\bibitem{Choi:1996fs}
K.~Choi, H.B.~Kim and J.E.~Kim, \emph{{Axion cosmology with a stronger QCD in
  the early universe}},
  \href{https://doi.org/10.1016/S0550-3213(97)00066-7}{\emph{Nucl. Phys. B}
  {\bfseries 490} (1997) 349}
  [\href{https://arxiv.org/abs/hep-ph/9606372}{{\ttfamily hep-ph/9606372}}].

\bibitem{Alonso-Alvarez:2023wig}
G.~Alonso-\'Alvarez, J.M.~Cline and T.~Xiao, \emph{{The Flavor of QCD Axion
  Dark Matter}},  \href{https://arxiv.org/abs/2305.00018}{{\ttfamily
  2305.00018}}.

\bibitem{DiLuzio:2015oha}
L.~Di~Luzio, R.~Gr\"ober, J.F.~Kamenik and M.~Nardecchia, \emph{{Accidental
  matter at the LHC}},
  \href{https://doi.org/10.1007/JHEP07(2015)074}{\emph{JHEP} {\bfseries 07}
  (2015) 074} [\href{https://arxiv.org/abs/1504.00359}{{\ttfamily
  1504.00359}}].

\bibitem{Borsanyi:2016ksw}
S.~Borsanyi et~al., \emph{{Calculation of the axion mass based on
  high-temperature lattice quantum chromodynamics}},
  \href{https://doi.org/10.1038/nature20115}{\emph{Nature} {\bfseries 539}
  (2016) 69} [\href{https://arxiv.org/abs/1606.07494}{{\ttfamily 1606.07494}}].

\bibitem{DEramo:2021lgb}
F.~D'Eramo, F.~Hajkarim and S.~Yun, \emph{{Thermal QCD Axions across
  Thresholds}}, \href{https://doi.org/10.1007/JHEP10(2021)224}{\emph{JHEP}
  {\bfseries 10} (2021) 224}
  [\href{https://arxiv.org/abs/2108.05371}{{\ttfamily 2108.05371}}].

\bibitem{Giudice:2000ex}
G.F.~Giudice, E.W.~Kolb and A.~Riotto, \emph{{Largest temperature of the
  radiation era and its cosmological implications}},
  \href{https://doi.org/10.1103/PhysRevD.64.023508}{\emph{Phys. Rev. D}
  {\bfseries 64} (2001) 023508}
  [\href{https://arxiv.org/abs/hep-ph/0005123}{{\ttfamily hep-ph/0005123}}].

\bibitem{Karamitros:2021nxi}
D.~Karamitros, \emph{{MiMeS: Misalignment mechanism solver}},
  \href{https://doi.org/10.1016/j.cpc.2022.108311}{\emph{Comput. Phys. Commun.}
  {\bfseries 275} (2022) 108311}
  [\href{https://arxiv.org/abs/2110.12253}{{\ttfamily 2110.12253}}].

\bibitem{CMB-HD:2022bsz}
{\scshape CMB-HD} collaboration, \emph{{Snowmass2021 CMB-HD White Paper}},
  \href{https://arxiv.org/abs/2203.05728}{{\ttfamily 2203.05728}}.

\bibitem{Stern:2016bbw}
I.~Stern, \emph{{ADMX Status}},
  \href{https://doi.org/10.22323/1.282.0198}{\emph{PoS} {\bfseries ICHEP2016}
  (2016) 198} [\href{https://arxiv.org/abs/1612.08296}{{\ttfamily
  1612.08296}}].

\bibitem{Salemi:2021gck}
C.P.~Salemi et~al., \emph{{Search for Low-Mass Axion Dark Matter with
  ABRACADABRA-10~cm}},
  \href{https://doi.org/10.1103/PhysRevLett.127.081801}{\emph{Phys. Rev. Lett.}
  {\bfseries 127} (2021) 081801}
  [\href{https://arxiv.org/abs/2102.06722}{{\ttfamily 2102.06722}}].

\bibitem{Alesini:2017ifp}
D.~Alesini, D.~Babusci, D.~Di~Gioacchino, C.~Gatti, G.~Lamanna and C.~Ligi,
  \emph{{The KLASH Proposal}},
  \href{https://arxiv.org/abs/1707.06010}{{\ttfamily 1707.06010}}.

\bibitem{DMRadio:2022pkf}
{\scshape DMRadio} collaboration, \emph{{Projected sensitivity of DMRadio-m3: A
  search for the QCD axion below 1\,\,\ensuremath{\mu}eV}},
  \href{https://doi.org/10.1103/PhysRevD.106.103008}{\emph{Phys. Rev. D}
  {\bfseries 106} (2022) 103008}
  [\href{https://arxiv.org/abs/2204.13781}{{\ttfamily 2204.13781}}].

\bibitem{Berlin:2020vrk}
A.~Berlin, R.T.~D'Agnolo, S.A.R.~Ellis and K.~Zhou, \emph{{Heterodyne broadband
  detection of axion dark matter}},
  \href{https://doi.org/10.1103/PhysRevD.104.L111701}{\emph{Phys. Rev. D}
  {\bfseries 104} (2021) L111701}
  [\href{https://arxiv.org/abs/2007.15656}{{\ttfamily 2007.15656}}].

\bibitem{deSalas:2016ztq}
P.F.~de~Salas and S.~Pastor, \emph{{Relic neutrino decoupling with flavour
  oscillations revisited}},
  \href{https://doi.org/10.1088/1475-7516/2016/07/051}{\emph{JCAP} {\bfseries
  07} (2016) 051} [\href{https://arxiv.org/abs/1606.06986}{{\ttfamily
  1606.06986}}].

\bibitem{Graf:2010tv}
P.~Graf and F.D.~Steffen, \emph{{Thermal axion production in the primordial
  quark-gluon plasma}},
  \href{https://doi.org/10.1103/PhysRevD.83.075011}{\emph{Phys. Rev. D}
  {\bfseries 83} (2011) 075011}
  [\href{https://arxiv.org/abs/1008.4528}{{\ttfamily 1008.4528}}].

\bibitem{Salvio:2013iaa}
A.~Salvio, A.~Strumia and W.~Xue, \emph{{Thermal axion production}},
  \href{https://doi.org/10.1088/1475-7516/2014/01/011}{\emph{JCAP} {\bfseries
  01} (2014) 011} [\href{https://arxiv.org/abs/1310.6982}{{\ttfamily
  1310.6982}}].

\bibitem{Arias:2023wyg}
P.~Arias, N.~Bernal, J.K.~Osi\'nski, L.~Roszkowski and M.~Venegas,
  \emph{{Revisiting signatures of thermal axions in nonstandard cosmologies}},
  \href{https://arxiv.org/abs/2308.01352}{{\ttfamily 2308.01352}}.

\bibitem{Hall:2009bx}
L.J.~Hall, K.~Jedamzik, J.~March-Russell and S.M.~West, \emph{{Freeze-In
  Production of FIMP Dark Matter}},
  \href{https://doi.org/10.1007/JHEP03(2010)080}{\emph{JHEP} {\bfseries 03}
  (2010) 080} [\href{https://arxiv.org/abs/0911.1120}{{\ttfamily 0911.1120}}].

\bibitem{DEramo:2017ecx}
F.~D'Eramo, N.~Fernandez and S.~Profumo, \emph{{Dark Matter Freeze-in
  Production in Fast-Expanding Universes}},
  \href{https://doi.org/10.1088/1475-7516/2018/02/046}{\emph{JCAP} {\bfseries
  02} (2018) 046} [\href{https://arxiv.org/abs/1712.07453}{{\ttfamily
  1712.07453}}].

\bibitem{Du:2021jcj}
Y.~Du, F.~Huang, H.-L.~Li, Y.-Z.~Li and J.-H.~Yu, \emph{{Revisiting dark matter
  freeze-in and freeze-out through phase-space distribution}},
  \href{https://doi.org/10.1088/1475-7516/2022/04/012}{\emph{JCAP} {\bfseries
  04} (2022) 012} [\href{https://arxiv.org/abs/2111.01267}{{\ttfamily
  2111.01267}}].

\bibitem{DEramo:2021psx}
F.~D'Eramo, F.~Hajkarim and S.~Yun, \emph{{Thermal Axion Production at Low
  Temperatures: A Smooth Treatment of the QCD Phase Transition}},
  \href{https://doi.org/10.1103/PhysRevLett.128.152001}{\emph{Phys. Rev. Lett.}
  {\bfseries 128} (2022) 152001}
  [\href{https://arxiv.org/abs/2108.04259}{{\ttfamily 2108.04259}}].

\bibitem{Brockway:1996yr}
J.W.~Brockway, E.D.~Carlson and G.G.~Raffelt, \emph{{SN1987A gamma-ray limits
  on the conversion of pseudoscalars}},
  \href{https://doi.org/10.1016/0370-2693(96)00778-2}{\emph{Phys. Lett. B}
  {\bfseries 383} (1996) 439}
  [\href{https://arxiv.org/abs/astro-ph/9605197}{{\ttfamily
  astro-ph/9605197}}].

\bibitem{Grifols:1996id}
J.A.~Grifols, E.~Masso and R.~Toldra, \emph{{Gamma-rays from SN1987A due to
  pseudoscalar conversion}},
  \href{https://doi.org/10.1103/PhysRevLett.77.2372}{\emph{Phys. Rev. Lett.}
  {\bfseries 77} (1996) 2372}
  [\href{https://arxiv.org/abs/astro-ph/9606028}{{\ttfamily
  astro-ph/9606028}}].

\bibitem{Payez:2014xsa}
A.~Payez, C.~Evoli, T.~Fischer, M.~Giannotti, A.~Mirizzi and A.~Ringwald,
  \emph{{Revisiting the SN1987A gamma-ray limit on ultralight axion-like
  particles}}, \href{https://doi.org/10.1088/1475-7516/2015/02/006}{\emph{JCAP}
  {\bfseries 02} (2015) 006} [\href{https://arxiv.org/abs/1410.3747}{{\ttfamily
  1410.3747}}].

\bibitem{Hamaguchi:2018oqw}
K.~Hamaguchi, N.~Nagata, K.~Yanagi and J.~Zheng, \emph{{Limit on the Axion
  Decay Constant from the Cooling Neutron Star in Cassiopeia A}},
  \href{https://doi.org/10.1103/PhysRevD.98.103015}{\emph{Phys. Rev. D}
  {\bfseries 98} (2018) 103015}
  [\href{https://arxiv.org/abs/1806.07151}{{\ttfamily 1806.07151}}].

\bibitem{Leinson:2021ety}
L.B.~Leinson, \emph{{Impact of axions on the Cassiopea A neutron star
  cooling}}, \href{https://doi.org/10.1088/1475-7516/2021/09/001}{\emph{JCAP}
  {\bfseries 09} (2021) 001}
  [\href{https://arxiv.org/abs/2105.14745}{{\ttfamily 2105.14745}}].

\bibitem{ciaran_o_hare_2020_3932430}
C.~O'HARE, \emph{cajohare/axionlimits: Axionlimits},  July, 2020.
\newblock 10.5281/zenodo.3932430.

\bibitem{Lee:2020cfj}
S.~Lee, S.~Ahn, J.~Choi, B.R.~Ko and Y.K.~Semertzidis, \emph{{Axion Dark Matter
  Search around 6.7 $\mu$eV}},
  \href{https://doi.org/10.1103/PhysRevLett.124.101802}{\emph{Phys. Rev. Lett.}
  {\bfseries 124} (2020) 101802}
  [\href{https://arxiv.org/abs/2001.05102}{{\ttfamily 2001.05102}}].

\bibitem{CAPP:2020utb}
{\scshape CAPP} collaboration, \emph{{First Results from an Axion Haloscope at
  CAPP around 10.7 $\mu$eV}},
  \href{https://doi.org/10.1103/PhysRevLett.126.191802}{\emph{Phys. Rev. Lett.}
  {\bfseries 126} (2021) 191802}
  [\href{https://arxiv.org/abs/2012.10764}{{\ttfamily 2012.10764}}].

\bibitem{McAllister:2017lkb}
B.T.~McAllister, G.~Flower, J.~Kruger, E.N.~Ivanov, M.~Goryachev, J.~Bourhill
  et~al., \emph{{The ORGAN Experiment: An axion haloscope above 15 GHz}},
  \href{https://doi.org/10.1016/j.dark.2017.09.010}{\emph{Phys. Dark Univ.}
  {\bfseries 18} (2017) 67} [\href{https://arxiv.org/abs/1706.00209}{{\ttfamily
  1706.00209}}].

\bibitem{Melcon:2018dba}
A.A.~Melc\'on et~al., \emph{{Axion Searches with Microwave Filters: the RADES
  project}}, \href{https://doi.org/10.1088/1475-7516/2018/05/040}{\emph{JCAP}
  {\bfseries 05} (2018) 040}
  [\href{https://arxiv.org/abs/1803.01243}{{\ttfamily 1803.01243}}].

\bibitem{Alesini:2020vny}
D.~Alesini et~al., \emph{{Search for invisible axion dark matter of mass
  m$_a=43~\mu$eV with the QUAX--$a\gamma$ experiment}},
  \href{https://doi.org/10.1103/PhysRevD.103.102004}{\emph{Phys. Rev. D}
  {\bfseries 103} (2021) 102004}
  [\href{https://arxiv.org/abs/2012.09498}{{\ttfamily 2012.09498}}].

\bibitem{MADMAX:2019pub}
{\scshape MADMAX} collaboration, \emph{{A new experimental approach to probe
  QCD axion dark matter in the mass range above 40 $\mu$eV}},
  \href{https://doi.org/10.1140/epjc/s10052-019-6683-x}{\emph{Eur. Phys. J. C}
  {\bfseries 79} (2019) 186}
  [\href{https://arxiv.org/abs/1901.07401}{{\ttfamily 1901.07401}}].

\bibitem{Ahyoune:2023gfw}
S.~Ahyoune et~al., \emph{{A proposal for a low-frequency axion search in the
  1-2 $\mu$eV range and below with the BabyIAXO magnet}},
  \href{https://arxiv.org/abs/2306.17243}{{\ttfamily 2306.17243}}.

\bibitem{Tang:2023oid}
Z.~Tang et~al., \emph{{SRF Cavity Searches for Dark Photon Dark Matter: First
  Scan Results}},  \href{https://arxiv.org/abs/2305.09711}{{\ttfamily
  2305.09711}}.

\end{thebibliography}\endgroup

\end{document}